\documentclass[%
 preprint,
 amsmath,amssymb,
 aps, physrev,
]{revtex4-2}

\usepackage{tikz}
\usepackage{pgfplots} 
\pgfplotsset{compat=1.18}
\usetikzlibrary{3d}

\usepackage{graphicx}
\usepackage{dcolumn}
\usepackage{bm}

\newcommand{\mathsfbi}[1]{\boldsymbol{\mathsf{#1}}}
\usepackage{xcolor}

\usepackage[colorlinks=true,allcolors=blue]{hyperref}
\newcommand{\rev}[1]{\textcolor{black}{#1}}

\begin{document}


\title{\textbf{Band-Ensemble Spectral Proper Orthogonal Decomposition with Frequency Attribution} 
}%

\author{Jakob G.~R.~von Saldern}
\email{Contact author: j.vonsaldern@tu-berlin.de}
\affiliation{%
Department of Civil and Mechanical Engineering, California Institute of Technology, Pasadena, California 91125, USA
}
\affiliation{
Laboratory for Flow Instabilities and Dynamics, Technische Universität Berlin, Berlin 10623, Germany
}%

\author{Oliver T.~Schmidt}
\affiliation{
Department of Mechanical and Aerospace Engineering, University of California, San Diego, La Jolla, California 92093, USA
}%

\author{Philipp Godbersen}
\affiliation{
Institute of Aerodynamics and Flow Technology, German Aerospace Center (DLR), Göttingen 37073, Germany
}%

\author{J.~Moritz Reumschüssel}
\affiliation{
Department of Engineering, University of Cambridge, Cambridge CB2 1PZ, UK
}%
\affiliation{
Chair of Fluid Dynamics, 
Technische Universität Berlin, Berlin 10623, Germany
}%

\author{Tim Colonius}
\affiliation{
Department of Civil and Mechanical Engineering, California Institute of Technology, Pasadena, California 91125, USA
}%


\date{\today}

\begin{abstract}
This study presents band-ensemble Spectral Proper Orthogonal Decomposition (bSPOD). The approach is inspired by frequency smoothing, a method used to reduce estimator variance in power spectral density estimates, and is here extended to SPOD. The algorithm estimates SPOD modes from consecutive Fourier coefficients obtained from a single Fourier transform of the full time record and thus avoids time segmentation. In this study, bSPOD is applied to artificial test data and to a PIV data set of a broadband–tonal cavity flow. Compared to the more commonly used Welch-based SPOD formulation, bSPOD reduces spectral leakage, permits increased frequency resolution, and retains frequency information of tonal components at comparable computational cost. These features enable reduced estimator variance while maintaining low bias for tonal components, making bSPOD particularly effective for broadband–tonal flows.  
\end{abstract}

\maketitle


\section{Introduction}\label{sec:Introduction}
Turbulent flow fields are inherently high-dimensional and chaotic.
Identifying organized spatio-temporal patterns within these chaotic dynamics, has long been central in fluid mechanics~\cite{lumley1967,sirovich1987turbulence,ROWLEY2009,SCHMID_2010,Taira2017,towne2018}. These so-called coherent structures are typically associated with high energy fluctuations and often underlie undesirable phenomena such as sound generation~\cite{Jordan2013,demange2024}, stall cell formation on airfoils~\cite{Sarras_2024}, combustion unsteadiness~\cite{Poinsot1987,vonSaldern25}, and structural vibration~\cite{Williamson2004}.

Techniques for processing such chaotic data have evolved over the last century. A central tool in this context is the power spectral density (PSD) which describes how a signal’s power is distributed over temporal frequency. 
In fluid mechanics, PSDs are widely used for processing velocity-probe and pressure-tap signals. The PSD is the Fourier transform of the autocorrelation function but is typically estimated from periodograms~\cite{wiener1930,Khintchine}. Because turbulence is chaotic, averaging is used to reduce estimator variance. 
\rev{For an estimate of the PSD, \rev{$\hat{S}(\omega)$}, this variance can be written as
\begin{equation}
\mathrm{Var}[\hat{S}(\omega)] =
\mathbb{E}\left[\left(\hat{S}(\omega)-\mathbb{E}[\hat{S}(\omega)]\right)^2\right],
\end{equation}
and quantifies the deviation of a finite-sample spectral estimate from its theoretical value. Since the latter is generally unknown a priori, the estimation variance cannot usually be evaluated directly. However, the estimation variance decreases with increasing number of statistically independent realizations employed in the averaging procedure. In practice, a high estimation variance manifests itself as noisy, jittering, or irregular spectral estimates. Consequently, the term estimation variance is often used more loosely to describe these statistical fluctuations, and is used in this sense throughout the present work.} Among averaging methods, Bartlett’s~\cite{bartlett1950} approach is the most commonly used: assuming ergodicity, the method divides the record into segments, computes a PSD for each, and averages the results. This reduces estimation variance at the cost of decreased frequency resolution, which is set by the segment length. \citet{welch1967} extended this approach by introducing overlapping segments and applying a taper to each segment to reduce spectral leakage.

\citet{Thomson1982} introduced the multitaper estimator, which multiplies the full time record by a set of orthogonal tapers to obtain approximately independent realizations. These tapered signals are Fourier transformed and averaged across tapers. This yields a PSD on the same frequency grid as the full-record DFT, but with a lower effective resolution because the estimate equals the true spectrum convolved with a spectral window formed from the tapers’ Fourier transforms~\cite{Babadi2014}. At matched effective resolution, multitaper generally reduces spectral leakage and provides comparable or lower variance than segment-based estimates under standard Gaussian assumptions~\cite{Bronez1992}. The trade-off is computational cost, since a full-record transform is computed for each taper.

A related approach to variance reduction is frequency smoothing~\cite{Daniell1946,Bartlett1948,Blackman1958}. \citet{Daniell1946} first proposed averaging periodogram values over adjacent frequencies, and \citet{Bartlett1948} noted the connection to truncating the autocorrelation. \citet{Blackman1958} then formalized the lag-window autocorrelation estimator by tapering the autocorrelation at large lags. The frequency-smoothing and lag-window views are equivalent when the smoothing filter and the lag window are a Fourier pair. In simple terms, this approach reduces estimator variance by averaging a full-record periodogram over neighboring frequencies, at the cost of lower effective frequency resolution.

Time segmentation, multitapering and frequency smoothing all reduce estimator variance but increase bias\rev{, i.e., the deviation between the expected spectral estimate and the true spectrum}, because finite-window operations either in time or frequency domain result in a spread of power across neighboring frequency bins. Where the PSD varies rapidly, for example at tonal peaks in the spectrum, stronger averaging broadens the narrow spectral features. A key distinction between the three approaches is that time-segmentation first splits the record into shorter segments, computes periodograms for each, and averages them; the effective frequency resolution is set by the segment length and is essentially uniform across the spectrum. In contrast, multitapering allows the number of tapers and their weights, which together determine the effective frequency resolution, to vary with frequency. Similarly, in frequency smoothing a single full-record periodogram is computed and then averaged across frequency, so the smoothing filter and its bandwidth can be chosen frequency by frequency, enabling a local bias–variance trade-off.

In fluid mechanics, not only temporal but also spatially resolved data are often available, for example, from particle image velocimetry (PIV) or from arrays of pressure or velocity probes sampled simultaneously. Consequently, in addition to temporal statistics one can evaluate spatial statistics to extract coherent structures. Spatial structures can be identified with proper orthogonal decomposition (POD), which, via an eigenvalue decomposition of the two-point spatial covariance, yields \rev{orthogonal spatial modes that optimally capture the flow energy}. The method follows the Karhunen–Loève framework and is also known as principal component analysis~\cite{Pearson01111901,hotelling1933}; it was introduced to fluid mechanics by \citet{lumley1967} and has since been widely used and extended. Although POD is not restricted to spatial coherence, it was initially used mainly to extract spatial structures~\cite{berkooz1993,Weiss2019}; To distinguish this most common variant of POD from others, the term space-only POD is also commonly used~\cite{towne2018}.

\citet{sirovich1987turbulence} introduced the method of snapshots, which reduces the cost of computing POD by solving an eigenproblem for the temporal correlation (snapshot) matrix that shares the non-zero eigenvalues with the spatial covariance matrix. \citet{boree2003} proposed extended POD to identify, in a second variable, the patterns most correlated with a reference measurement. \citet{Sieber2016} applied a time-filtered POD, yielding modes whose spectra are concentrated within a frequency band. \citet{Mendez2019} combined multiresolution (wavelet) analysis with POD; the resulting multiscale POD isolates structures across scales and is not restricted to statistically stationary processes. The natural extension to this is space–time POD, which identifies an energy-optimal basis for the fully space- and time-resolved signal, without \textit{a priori} scale restrictions~\cite{lumley1967,gordeyev2013,Frame2023}. This, however, requires an eigenvalue decomposition of the space–time \rev{covariance} and is computationally expensive. Modified versions that introduce temporal conditioning to predict extreme events~\cite{Schmidt_Schmid_2019} and generate forecasts~\cite{Schmidt_RSPA_2026} have been introduced more recently.
 
Many relevant flows are statistically stationary, so space–time POD is unnecessary, and energy-optimal spatial modes can be identified as a function of frequency~\cite{lumley1967,lumley1970}. These are obtained by an eigendecomposition of the cross-spectral-density (CSD) matrix, a method first termed spectral proper orthogonal decomposition (SPOD) by \citet{picard2000}. SPOD identifies an energy-optimal basis at each frequency~\cite{lumley1967}. 
\rev{Its application} relies on large datasets to estimate the CSD across frequencies. 
In practice, Welch segmentation is used: the record is assumed to be ergodic, split into blocks, Fourier modes are computed per block, and the CSD is \rev{estimated} from the ensemble of blockwise realizations~\cite{Schmidt2020}.
SPOD is widely applied~\cite{picard2000,CITRINITI2000,Chen2024,vonSaldern24,Steinfurth2025}, also due to its link to resolvent analysis~\cite{towne2018}, which provides a physics-based model for the data-driven modes. Variants and extensions of SPOD have also been proposed. \citet{Colanera2025} introduced a robust POD formulation to mitigate sensitivity to outliers, improving mode convergence on noisy datasets. \citet{Blanco2022} applied a time shift to align advecting structures and thereby improve SPOD mode convergence. \citet{Heidt2024a} extended SPOD to cyclostationary flows in which the broadband statistics are periodic rather than stationary.

Despite its success in identifying temporally and spatially coherent structures, SPOD with Welch averaging inherits the \rev{estimation} variance–bias trade-off of spectral estimation from finite records~\cite{towne2018,yeung2024}. Increasing frequency resolution (longer segments) improves the accuracy of frequency and power estimates for sharp spectral features, but yields high variance and slow convergence in broadband regions. Increasing the number of blocks (shorter segments) reduces variance and improves convergence, but degrades frequency resolution and biases estimates near sharp features. To improve this trade-off, \citet{schmidt2022multi} introduced multitaper SPOD, leveraging Thomson's~\cite{Thomson1982} multitaper estimator. As noted above, multitapering still entails a bias–variance trade-off, but it outperforms Welch segmentation for PSD estimation at matched effective resolution~\cite{Bronez1992}. To reduce the significant increase in computational cost associated with multitapering, Schmidt~\cite{schmidt2022multi} proposed a hybrid SPOD that combines Welch segmentation with multitapering. Building on this, \citet{yeung2024} developed an adaptive SPOD formulation inspired by \citet{Riedel1995}, based on sinusoidal tapers. The tapered Fourier-mode estimates are obtained from a single Fourier transform of the signal, using zero padding to twice the record length, which substantially reduces cost. Following the idea of adapting the estimator with frequency~\cite{Riedel1995,BARBOUR2014}, the number of tapers is selected by an alignment–convergence criterion: broadband regions use more tapers to ensure convergence, whereas near spectral peaks, where tonal energy yields rapid convergence, fewer tapers are used to keep spectral bias low. 
In this way, the adaptive SPOD method \rev{improves convergence of} the \rev{spectral estimate}, reduces variance in broadband regions, and keeps the bias near tonal components low. However, the approach is iterative and may require many eigenvalue decompositions before convergence, which entails high computational cost.

In this study, we present an algorithm for computing SPOD modes based on ensembles of Fourier modes within a frequency band, referred to as band-ensemble SPOD (bSPOD). The idea is inspired by frequency smoothing, which reduces variance in PSD estimates by averaging a full-record periodogram over neighbouring frequencies. For standalone PSD estimation, Welch’s segmentation is more commonly used than frequency smoothing. However, when combined with POD in the band-ensemble strategy, the approach unfolds its full potential. 
The modifications to the commonly used SPOD algorithm with Welch's ansatz are small, yet the advantages are substantial: bSPOD reduces spectral leakage and provides a data-driven frequency estimate for each mode, enabling variance reduction in broadband regions while keeping bias near tonal eigenvalues low, without relying on an adaptive procedure. Computational cost is comparable to SPOD with Welch segmentation. Since the smoothing is applied after computing the spectrum, bSPOD allows the effective frequency resolution to be modified without requiring recomputation of the Fourier transform.

In the following, we first introduce standard Welch-based SPOD and then describe bSPOD. The methods are compared on an artificial dataset to highlight the advantages of bSPOD and on experimental PIV measurements of a broadband–tonal cavity flow to demonstrate applicability to real data. Results and features are discussed, and the study concludes with a summary of the main findings.

\section {Methodology}
In this section, we outline the Welch-based SPOD algorithm and its band-ensemble adaptation. We focus on the time-discrete formulation and refer the reader to the literature for further details \cite{lumley1970,towne2018,Schmidt2020,yeung2024}.

\rev{A key focus of the present study is broadband–tonal flow configurations. SPOD estimates a CSD under stationarity assumptions. A true spectral line corresponds to infinite temporal coherence and is therefore fundamentally singular relative to the broadband component. In finite data, the distinction between a line spectrum, a narrowband stochastic process, and cyclostationary dynamics is not identifiable via the Fourier transform.  Accordingly, the present methods should be interpreted operationally as a decomposition for finite broadband–tonal datasets, rather than as evidence that the underlying process admits a strictly stationary ergodic spectral representation}

\subsection {Spectral Proper Orthogonal Decomposition}\label{subsec:SPOD}
We consider a discrete space–time signal 
$q(x_l,t_n)$, with $l=0,\dots,N_x-1$ spatial degrees of freedom and $n=0,\dots,N_t-1$ time samples. Here, we use space and time as coordinates as those are the most common encountered in fluid dynamics. Multiple spatial directions and physical variables can be stacked into the $l$-index.
The time series is divided into $N_b$ blocks of length $N_w$. To increase the number of blocks, adjacent segments may be allowed to overlap. In the following, we focus on the core algorithm without overlapping and consider overlapping separately in Section~\ref{sec:overlap}. Each block is transformed into frequency space using a temporal discrete Fourier transform,
\begin{equation}
\hat{q}^{(b)}(x_l,f_j) = \frac{1}{N_w}\sum_{n=0}^{N_w-1} w(t_n) q^{(b)}(x_l,t_n)\,
\exp\!\left(- i \,2\pi j n /N_w \right),
\qquad j=0,\dots,N_w-1, \label{eq:dft}
\end{equation}
where $i$ denotes the imaginary unit, $b$ the block index, $j$ the frequency index, and \rev{$w(t_n)$} a taper (window) used to reduce spectral leakage. The hat symbol denotes a complex Fourier amplitude. The frequency \rev{spacing} results from the length of the windows, $\Delta f = 1/(N_w \Delta t)$, where $\Delta t$ denotes the time step between two consecutive time samples.
For each frequency bin, the Fourier coefficients from the different blocks can be arranged into the data matrix
\begin{equation}
\mathsfbi{Q}_j = \big[\, \hat{\mathbf{q}}^{(1)}_j,\,\hat{\mathbf{q}}^{(2)}_j,\,\dots,\,\hat{\mathbf{q}}^{(N_b)}_j \,\big], \label{eq:Qwelch}
\end{equation}
where each column $\hat{\mathbf{q}}^{(b)}_j$ is a spatially resolved Fourier coefficient at frequency $j$ associated with block $b$.
The shape of $\mathsfbi{Q}_j$ is $N_x \times N_b$. By averaging over all blocks an estimate of the cross-spectral density (CSD) matrix at frequency $j$ is obtained
\begin{equation}
\mathsfbi{C}_j = \frac{\alpha}{N_b}\, \mathsfbi{Q}_j \mathsfbi{Q}_j^{*}, \label{eq:CSDwelch}
\end{equation}
where $(\cdot)^{*}$ denotes the conjugate transpose. The prefactor is
\begin{equation}
    \alpha = \frac{1}{\Delta f \, U}, 
    \qquad U=\frac{1}{N_w}\sum_{n=0}^{N_w-1} w^2(t_n), 
\end{equation}
which normalizes the window energy and ensures that the SPOD eigenvalues have units of power spectral density (variance per unit frequency). The definition of $\alpha$ is consistent with the forward transform normalization in Eq.~\eqref{eq:dft} and ensures that Parseval’s theorem is respected~\cite{yeung2024}.

An eigenvalue decomposition of $\mathsfbi{C}_j$ yields the SPOD modes
\begin{equation}
    \mathsfbi{C}_j \mathsfbi{W} \, \boldsymbol{\Phi}_{j} = \boldsymbol{\Lambda}_{j}\, \boldsymbol{\Phi}_{j},
\end{equation}
where $\mathsfbi{W}$ is a weight matrix that defines the discretized inner product. The eigenvectors contained in the columns of $\boldsymbol{\Phi}$ are the SPOD modes that are orthonormal with respect to the inner product, $\boldsymbol{\Phi}^* \mathsfbi{W}\boldsymbol{\Phi} = \mathsfbi{I}$. The spectral power density of the modes is contained in the diagonal matrix $\boldsymbol{\Lambda}$. If the number of spatial degrees of freedom is larger than the number of blocks, one typically solves an alternative eigenvalue problem with the same non-zero eigenvalues known as the method of snapshots~\cite{sirovich1987turbulence}
\begin{equation}
\frac{\alpha}{N_b}\mathsfbi{Q}_j^* \mathsfbi{W} \mathsfbi{Q}_j \boldsymbol{\Theta}_{j} = \boldsymbol{\Lambda}_{j}\, \boldsymbol{\Theta}_{j},
\end{equation}
where $\boldsymbol{\Theta}_{j}$ are the expansion coefficients that in space-only POD are the time coefficients. This eigenvalue problem is of size $N_b \times N_b$. The SPOD modes follow from
\begin{equation}\label{eq:SPOD_expansion_method_of_snaps}
    \boldsymbol{\Phi}_{j} = \sqrt{\frac{\alpha}{N_b}} \mathsfbi{Q}_j  \boldsymbol{\Theta}_{j} \boldsymbol{\Lambda}_{j}^{-\frac{1}{2}}.
\end{equation}
This expression shows that the expansion coefficients $\boldsymbol{\Theta}_{j}$ in SPOD quantify how much each block-Fourier coefficient contained in $\mathsfbi{Q}_j$ contributes to a given SPOD mode in $\boldsymbol{\Phi}_{j}$.

Overall this yields $N_b$ SPOD modes that can be ranked according to the power density contained in the eigenvalue. All modes are associated with the respective Fourier-bin frequency $f_j = j \,\Delta f$, with
\begin{equation}\label{eq:fourier_bins}
  j =
  \begin{cases}
    -N_w/2,\dots,+N_w/2-1, & N_w \ \text{even}, \\[6pt]
    -(N_w-1)/2,\dots,+ (N_w-1)/2, & N_w \ \text{odd}.
  \end{cases}
\end{equation}
Compared to the definition in Eq.~\ref{eq:dft}, the frequency index is shifted such that zero frequency is centered.
Since the window length $N_w$ is determined by the total duration of the time series $N_t$ and the number of blocks $N_b$, SPOD faces a trade-off between frequency resolution and statistical convergence. Increasing the frequency resolution yields fewer Fourier coefficient samples, and therefore less well-converged eigenvalues, which manifests as higher variance in the estimated modal power spectral density. Conversely, increasing the number of samples by using shorter windows improves statistical convergence but reduces frequency resolution and increases spectral leakage. This results in a loss of spectral sharpness, commonly referred to as spectral bias, which is particularly pronounced when distinct tonal components are smeared \rev{into neighboring frequency bins due to leakage}. This trade-off between spectral variance and spectral bias is a well known problem in spectral estimation.

When considering this trade-off it is important to recall that the SPOD modes form an energy-optimal basis for the dynamics \rev{of each frequency bin. Owing to the finite segment length, each bin receives contributions from a range of frequencies through spectral leakage, with the strongest contributions originating from neighboring, unresolved spectral content. 
One approach to quantify the effective width of this most contributing frequency band is the equivalent noise bandwidth, which for a rectangular window is equal to $\Delta f$ \citep{harris1978}.
\rev{Consequently, the additional samples obtained by considering shorter segments only improve the convergence of the spectral estimate if the flow dynamics contributing within the widened effective bandwidth $\Delta f$ remain sufficiently similar.} This requires that the flow dynamics do not vary abruptly with frequency.}

\rev{We note that tapering decreases spectral leakage by reducing the amplitude of side lobes of the spectral window function at the cost of increasing the main lobe width, which in turn increases the effective bandwidth~\citep{harris1978}. In simple terms, tapering reduces long-range leakage but increases local spectral smearing.}


\subsection{Band-ensemble Spectral Proper Orthogonal Decomposition}
The band-ensemble SPOD (bSPOD) builds on the frequency-smoothing approach used in spectral estimation that is extended to time–space data. In contrast to the Welch-based algorithm, bSPOD starts with a decisive difference: the data are not segmented into temporal blocks, but the discrete Fourier transform is computed 
over the entire time series,
\begin{equation}
\tilde{q}(x_l,f_\gamma) = \frac{1}{N_t}\sum_{n=0}^{N_t-1} q(x_l,t_n)\,
\exp\!\left(- i \,2\pi \gamma n /N_t \right),
\qquad \gamma=0,\dots,N_t-1.
\end{equation}
To distinguish symbols from the Welch-based approach, we use a tilde.
The frequency \rev{spacing} of the Fourier transform, $\widetilde{\Delta f} = 1/(N_t \Delta t)$, is thus increased by the  ratio of total to block length, i.e.\ $N_t/N_w$, compared to the standard Welch-based SPOD. \rev{No tapering is applied in the present formulation. While tapering is not fundamentally incompatible with the method, bSPOD relies on a single Fourier transform of the full-length record, for which the increased record length yields a narrower spectral window associated with the implicitly assumed rectangular window and thus naturally reduces spectral leakage compared to the shorter segments used in Welch-based SPOD. Tapering can further suppress the side lobes of the spectral window, but at the cost of broadening its main lobe and thereby increasing correlations between neighboring Fourier coefficients, as already mentioned above.}

In contrast to the Welch approach, the data matrix is not constructed from repeated realizations at a fixed frequency, but from $N_f$ consecutive Fourier modes,
\begin{equation}
\Tilde{\mathsfbi{Q}}_j = \big[\, \tilde{\mathbf{q}}_{j+1},\,\tilde{\mathbf{q}}_{j+2},\,\dots,\,\tilde{\mathbf{q}}_{j+N_f} \,\big], 
\qquad j = b N_f,\quad b=0,\dots,N_t/N_f, \label{eq:Qsmoothing}
\end{equation}
where $\tilde{\mathbf{q}}_\gamma$ denotes the full length Fourier mode at frequency $\gamma$.
To illustrate the connection with the Welch-based formulation, we consider the case where the number of blocks in the Welch formulation equals the number of consecutive Fourier modes considered in the band-ensemble ansatz, $N_b = N_f$, and no overlap is used. In this situation, the Welch-SPOD matrix $\mathsfbi{Q}_j$ contains $N_b$ realizations of a Fourier mode defined on \rev{a} frequency \rev{grid with spacing} $\Delta f$, while the band-ensemble matrix $\tilde{\mathsfbi{Q}}_j$ contains $N_f$ consecutive Fourier modes at an enhanced resolution of $\widetilde{\Delta f} = \Delta f/N_f$. The corresponding analogue of the cross-spectral density matrix can be approximated as
\begin{equation}
    \tilde{\mathsfbi{C}}_j \approx \frac{\tilde{\alpha}}{N_f} \,\tilde{\mathsfbi{Q}}_j \tilde{\mathsfbi{Q}}_j^{*},\label{eq:CSDb}
\end{equation}
where $\tilde{\alpha} = 1/\widetilde{\Delta f}$ is adopted to the increased frequency resolution in the Fourier transform step. With this definition of the prefactors, choosing $ N_f = N_b $ implies that both methods consider the dynamics \rev{with} the same frequency \rev{spacing}
$\Delta f = N_f \widetilde{\Delta f}$, and use the same number of samples per \rev{estimate}. This can be seen by comparing Eqs.~\eqref{eq:CSDwelch} and \eqref{eq:CSDb}:
Without overlap and for  $ N_f = N_b $, the CSD estimate in Welch-SPOD can be interpreted as the average over $N_b$ realizations, each defined \rev{with an effective bandwidth} of $\Delta f$ (see Eq.~\eqref{eq:CSDwelch}); the prefactor becomes $\Delta f/N_b$ assuming no tapering is applied ($U=1$). 
In bSPOD, under the same conditions, the prefactor $\tilde{\alpha}/N_f$ in the CSD definition (Eq.~\eqref{eq:CSDb}) reduces to $\Delta f$ such that the resulting matrix represents the cross-spectral density \rev{with} the same \rev{effective} frequency \rev{bandwidth}.
The remaining factor $1/N_b$ differs between the two methods because bSPOD does not average $N_b$ independent realizations, each defined \rev{with frequency spacing} $\Delta f$, as is done in SPOD with Welch segmentation. Instead, it aggregates the contributions of $N_f\, (=N_b)$ consecutive Fourier modes within the band $\Delta f$, each defined on an $N_f$-times finer resolution  $\widetilde{\Delta f} = \Delta f / N_f$. Consequently, the prefactor ensures that both CSD matrices represent the same total power \rev{within the effective bandwidth} $\Delta f$. At the same time, estimating the CSD from consecutive Fourier modes at higher frequency resolution is advantageous because in-band information can be recovered through the expansion coefficients, as shown in the following.

Due to the importance of the expansion coefficients, bSPOD relies on applying the method of snapshots
\begin{equation}\label{eq:bSPOD_m_of_snaps}
    \frac{\tilde{\alpha}}{N_f} \,\tilde{\mathsfbi{Q}}_j^{*}\,\mathsfbi{W}\,\tilde{\mathsfbi{Q}}_j \,\tilde{\boldsymbol{\Theta}}_{j}
    = \tilde{\boldsymbol{\Lambda}}_{j}\,\tilde{\boldsymbol{\Theta}}_{j},
\end{equation}
which is of size $N_f \times N_f$. The bSPOD modes then follow as
\begin{equation}\label{eq:bSPOD_n_of_snaps2}
    \tilde{\boldsymbol{\Phi}}_{j} = 
    \sqrt{\frac{\tilde{\alpha}}{N_f}}\;\tilde{\mathsfbi{Q}}_j\,\tilde{\boldsymbol{\Theta}}_{j}\,
    \tilde{\boldsymbol{\Lambda}}_{j}^{-1/2},
\end{equation}
so that $\tilde{\boldsymbol{\Phi}}_{j}^{*}\mathsfbi{W}\tilde{\boldsymbol{\Phi}}_{j}
= \mathsfbi{I}$. 
Analogous to the method of snapshot in Welch-based SPOD, the expansion coefficient in Eq.~\eqref{eq:bSPOD_n_of_snaps2} quantify how much each element of $\tilde{\mathsfbi{Q}}_j$ contributes to each bSPOD mode contained in $\tilde{\boldsymbol{\Phi}}_{j}$. Because in bSPOD each element in $\tilde{\mathsfbi{Q}}_j$ is a Fourier mode defined on a narrow and unique frequency bin, this information can be exploited to obtain in-band frequency information for each mode.
Specifically, we assign each bSPOD mode a representative, data-driven frequency by using the expansion coefficients as frequency weights. The columns of $\tilde{\boldsymbol{\Theta}}_{j}$
contain the contribution of each discrete Fourier mode to a given bSPOD mode; we therefor define frequency weights as
\begin{equation}
    \beta_{\eta,\gamma} = \frac{|{\theta}_{\gamma,\eta}|^{\ell}}
    {\sum_{\gamma^\prime=1}^{N_f} |{\theta}_{\gamma^\prime,\eta}|^{\ell}},
    \qquad \sum_{\gamma^\prime=1}^{N_f} \beta_{\eta,\gamma^\prime} = 1 ,\label{eq:weight}
\end{equation}
where ${\theta}_{\gamma,\eta}$ are the elements of $\tilde{\boldsymbol{\Theta}}_{j}$ and $\ell$ is an exponent to control the type of weighting.
The data-driven frequency estimates for the $N_f$ ranked bSPOD modes $\eta$ are then obtained by a matrix vector product between the weights $\boldsymbol{\beta}$ and the frequencies $\mathbf{f}_j$ of the discrete Fourier modes contained in $\tilde{\mathsfbi{Q}}_j$ (see Eq.~\eqref{eq:Qsmoothing}) 
\begin{equation}\label{eq:weighted_freqs}
    \tilde{\mathbf{f}}_j = \boldsymbol{\beta}\, \mathbf{f}_j 
    \qquad \mathbf{f}_j=\big[f_{j+1},\,f_{j+2},\,\ldots,\,f_{j+N_f}\big]^{\top}.
\end{equation}
Choosing $\ell = 0$, the weights become $1/N_f$ and each mode is attributed to the window center frequency. With $\ell = 1$, the average is weighted by the relative contributions of the Fourier modes to the bSPOD mode. For $\ell=\infty$, each bSPOD mode is assigned the frequency of its single most contributing Fourier mode.

Overall, the data-driven frequency attribution allows each bSPOD mode, though defined over a frequency band, to be associated with a more precise in-band frequency estimate. This is made possible by forming the CSD matrix from neighboring Fourier modes obtained from a single Fourier transform of the full time record. The key are the expansion coefficients that allow in-band frequency information of the high-resolution Fourier modes to be maintained, when the optimal basis is computed through POD of consecutive Fourier modes.

\rev{We emphasize that both Welch-based SPOD and bSPOD seek to approximate the same underlying CSD operator. Consequently, provided that the CSD approximation is sufficiently converged and the underlying assumptions hold, the resulting modes retain the theoretical properties associated with SPOD, including the established connection to resolvent analysis~\cite{towne2018}.}
An open-source MATLAB implementation of bSPOD is available online~\footnote{The code is available at \url{https://github.com/jakobvonsaldern/bspod}.}. 

\subsection{Band-ensemble filter length}
The main parameter that affects bSPOD is the frequency-window length $N_f$, which, as introduced above, plays a role analogous to the number of blocks, $N_b$, in Welch-SPOD. For small $N_f$, the bSPOD modes are computed from only a few consecutive Fourier modes, which improves frequency resolution but reduces statistical convergence, as the eigenvalue problem is small and the dynamics in the considered band must be represented by few modes. Increasing $N_f$ adds consecutive Fourier modes: the modes in total have to represent the dynamics over a wider band, but the larger problem also provides more flexibility to distribute power among the modes. If the frequency content changes slowly across frequency, convergence improves as more Fourier modes are included.
Particularly interesting are the limits $N_f=1$ and $N_f=N_t$. In the limiting case $N_f=1$, the data matrix $\tilde{\mathsfbi{Q}}_j$ reduces to a single column,
\begin{equation}
  \mathsfbi{Q}_j \;=\; \big[\, \tilde{\mathbf{q}}_j \, \big],
\end{equation}
so the corresponding eigenvalue problem becomes trivial and the bSPOD mode coincides with the normalized discrete Fourier mode itself, with its squared amplitude $\tilde{\mathbf{q}}_j^*\tilde{\mathbf{q}}_j$ as the associated eigenvalue. In the opposite limit $N_f=N_t$, all Fourier modes are included and the modes simplify to space-only POD modes. This can be shown by defining the mean-subtracted space–time data matrix
\begin{equation}
  \mathsfbi{X} \;=\; \big[\mathbf{q}_0, \, \mathbf{q}_1, \,..., \, \mathbf{q}_{N_t-1} \big]
  \;\in\;\mathbb{C}^{N_x\times N_t},
\end{equation}
with $N_x$ spatial and $N_t$ temporal degrees of freedom, and the unitary DFT matrix
\begin{equation}
  \mathsfbi{F}\in\mathbb{C}^{N_t\times N_t},\qquad
  F_{l,n} \;=\; \frac{1}{\sqrt{N_t}}\,
  e^{-\,i\,2\pi n l / N_t}, \qquad 
  \mathsfbi{F}^* \mathsfbi{F} \;=\; \mathsfbi{I}.
\end{equation}
This allows us to write the bSPOD data matrix as $ \tilde{\mathsfbi{Q}}_j \;=\; \mathsfbi{X}\,\mathsfbi{F}$. Hence the bSPOD cross-spectral density estimate, according to Eq.\eqref{eq:CSDb} becomes
\begin{equation}
  \tilde{\mathsfbi{C}}_j \;=\; \frac{\Delta t}{N_t}\,\tilde{\mathsfbi{Q}}_j\tilde{\mathsfbi{Q}}_j^{*}
  \;=\; \frac{\Delta t}{N_t}\,\mathsfbi{X}\,\mathsfbi{F}\mathsfbi{F}^{*}\,
  \mathsfbi{X}^{*}
  \;=\; \frac{\Delta t}{N_t}\,\mathsfbi{X}\mathsfbi{X}^{*},\label{eq:PODcov}
\end{equation}
where the prefactor varies slightly from the definition in Eq.\eqref{eq:CSDb} since the Fourier transform is based on a unitary normalization. After normalizing with the time step $\Delta t$, Eq.~\eqref{eq:PODcov} becomes the time-averaged covariance matrix. Thus the bSPOD eigenproblem reduces to the space-only POD covariance problem. From this we conclude that the parameter $N_f$ spans the transition from $N_f=1$, where the modes reduce to DFT modes and capture only temporal coherence, to $N_f=N_t$, where the modes become space-only POD modes that optimally represent the spatial second-order statistics of the entire time series. Intermediate values of $N_f$ yield bSPOD modes \rev{that optimally represent the spatial second-order statistics within a local frequency band}, and thus structures of temporal and spatial coherence. The same limits are reached for Welch-based SPOD: with a single block (no segmentation) the CSD estimate is rank-1 and the mode equals the normalized DFT mode; with $N_t$ blocks of length one, the CSD matrix becomes the covariance matrix, since the DFT of a single time sample is the sample itself, and the SPOD modes become POD modes. 
Thus, in these parameter limits, bSPOD converges to the same endpoints as other SPOD algorithms, including Welch-based SPOD and the time-filtered POD of~\citet{Sieber2016}.

A key advantage of bSPOD is that the band-ensemble filter length $N_f$ does not have to be constant across the spectrum, but may vary with frequency. In practice, shorter bands can be chosen in regions where modes are expected to undergo rapid changes in spatial wavelength, while larger bands can be used where the modes vary more slowly. 
This is especially relevant at low frequencies, where relative frequency changes are naturally larger. Since spatial wavelengths are often coupled to temporal frequency through a dispersion relation, this can lead to substantial variation in the spatial extent of coherent structures. For turbulent jets, for example, Welch-based SPOD can suffer from this effect, and variable frequency-bin widths have been recommended~\cite{heidt2024}. The advantage of bSPOD is that the band-ensemble filter length can be adapted locally in the spectrum without recomputing the Fourier transform\rev{, allowing the effective bSPOD frequency resolution to vary with frequency.} In contrast, the Welch-based SPOD algorithm requires recomputing the Fourier coefficients whenever the frequency resolution (block length) is changed, so modifying the frequency resolution entails complete reprocessing of the data.

\subsection{Summary of differences}
\begin{figure}
\begin{tikzpicture}[z={(-10:10mm)},x={(20:6mm)},scale=1.25,font=\footnotesize]

\newcommand{\drawColorGradient}[4]{%

  \pgfmathsetmacro{\steps}{#1}%
  \pgfmathsetmacro{\offset}{#4}%

  \foreach \step in {0,...,\steps}{%
    \pgfmathsetmacro{\percent}{\step/\steps*100}          
    \pgfmathsetmacro{\z}{\step/10+\offset}                
    \begin{scope}[canvas is xy plane at z=\z]
      \draw[fill=#3!\percent!#2] (0,0) rectangle (1,1);
    \end{scope}%
  }%
}

\newcommand{\drawColorGradientStack}[7]{%

  \pgfmathsetmacro{\steps}{#1}%
  \pgfmathsetmacro{\offset}{#4}%
  \pgfmathsetmacro{\maxz}{\steps/10+\offset}%

  \foreach \step in {0,...,\steps}{%
    \pgfmathsetmacro{\percent}{\step/\steps*100}%
    \pgfmathsetmacro{\z}{\step/10+\offset}%
    \begin{scope}[canvas is xy plane at z=\z]%
      \draw[fill=#3!\percent!#2] (0,0) rectangle (1,1);%
    \end{scope}%
  }%

  \draw[#5,#6] (0,0,\maxz) -- (0,1,\maxz) --(1,1,\maxz) -- (1,0,\maxz) -- (0,0,\maxz) -- (0,0,\offset) -- (0,1,\offset)-- (0,1,\maxz);
  \draw[#5,#6] (0,1,\offset) -- (1,1,\offset)-- (1,1,\maxz) node[black,pos=0.5,above right]{#7};
 }

\newcommand{\drawHueGradient}[5]{%
  \pgfmathsetmacro{\steps}{#1}
  \pgfmathsetmacro{\hStart}{#2}
  \pgfmathsetmacro{\hEnd}  {#3}
  \pgfmathsetmacro{\zOffset}{#4}
  \pgfmathsetmacro{\yOffset}{#5}
  %
  \foreach \step in {0,...,\steps}{%
    \pgfmathsetmacro{\h}{\hStart + (\hEnd-\hStart)*\step/\steps}
    \pgfmathsetmacro{\z}{\step/10 + \zOffset}
    %
    \pgfmathsetmacro{\hue}{360*\step/6}
    \definecolor{tmpcolor2}{Hsb}{\h,1,0.9};
    
    \begin{scope}[canvas is xy plane at z=\z]%
      \draw[tmpcolor2,fill=black!0,very thick]%
        (0,\yOffset) rectangle +(1,1);%

    \end{scope}
  }%
  };

   \tikzset{
        ellC/.style={
            color of colormap={#1},
            draw=black,fill=.,
            thick
        },
  }

  \newcommand{\drawColormapStack}[4]{%
  \pgfmathsetmacro{\steps}{#1}%
  \pgfmathsetmacro{\zOff}{#3}%
  \pgfmathsetmacro{\yOff}{#4}%
  \foreach \step in {0,...,\steps}{%
    \pgfmathsetmacro{\pct}{\step/\steps*1000}%
    \pgfmathsetmacro{\z}{\step/10 + \zOff}%
    \begin{scope}[canvas is xy plane at z=\z]%
      \draw[ellC={\pct of #2}] (0,\yOff) rectangle (1,1+\yOff);%
    \end{scope}%
  }%

}

  \newcommand{\drawColormapStackUni}[6]{%
  \pgfmathsetmacro{\steps}{#1}%
  \pgfmathsetmacro{\zOff}{#3}%
  \pgfmathsetmacro{\yOff}{#4}%
  \foreach \step in {0,...,\steps}{%
    \pgfmathsetmacro{\pct}{\step/\steps*1000}%
    \pgfmathsetmacro{\z}{\step/10 + \zOff}%
    \begin{scope}[canvas is xy plane at z=\z]%
      \draw[ellC={#5 of #2}] (0,\yOff) rectangle (1,1+\yOff);%
    \end{scope}%
  }%
  
  \pgfmathsetmacro{\z}{\steps/10 + \zOff}
    \begin{scope}[canvas is xy plane at z=\z]
      \path (-0.2,\yOff) rectangle (1,2+\yOff) node[pos=0.1,above right, gray!80]{${f}_{#6}$};
    \end{scope}
    
     \pgfmathsetmacro{\z}{\steps/10 /2 + \zOff}
    \begin{scope}[canvas is xy plane at z=\z]
      \path (0,\yOff) rectangle (1,1+\yOff) node[pos=-0.37,anchor=south west]{$\boldsymbol{\mathsf{Q}}_{#6}$};
    \end{scope}
 
}

\draw[thick,-stealth] (0,-0.1,0) -- (0,-0.1,3.5) node[pos=1,right]{$t$};

\drawColorGradientStack{5}{black!70!white}{black!40!white}{0}{green!50!black}{very thick}{$b=1$}
\drawColorGradientStack{5}{black!40!white}{black!10!white}{1}{orange!90!black}{very thick}{$b=2$}

\draw[dotted,thick] (0.5,0.5,1.9) -- (0.5,0.5,2.6);

    \begin{scope}[canvas is xy plane at z=1.7]
        \draw[thick,->] (0,0) -- (0,0.5) node[pos=1.2]{$y$};
        \draw[thick,->] (0,0) -- (0.5,0) node[pos=1.4]{$x$};
    \end{scope}

\drawColorGradientStack{5}{black!10!white}{black!0!white}{3}{blue!80!black}{very thick}{$b=N_b$}

\newcommand{\cmap}{viridis}

\drawColormapStack{5}{\cmap}{0}{-1.7};
\draw[thick,-stealth] (0,-1.8,0) -- (0,-1.8,0.7) node[pos=0.9,right]{$f$};

\drawColormapStack{5}{\cmap}{1}{-1.7};
\draw[dotted,thick] (0.5,0.5-1.7,1.9) -- (0.5,0.5-1.7,2.6);
\draw[thick,-stealth] (0,-1.8,1) -- (0,-1.8,1.7) node[pos=0.9,right]{$f$};

\drawColormapStack{5}{\cmap}{3}{-1.7}
\draw[thick,-stealth] (0,-1.8,3) -- (0,-1.8,3.7) node[pos=0.9,right]{$f$};

\drawColormapStackUni{5}{\cmap}{0}{-3.8}{0}{1}

\drawColormapStackUni{5}{\cmap}{1}{-3.8}{400}{2}
\draw[dotted,thick] (0.5,0.5-3.8,1.9) -- (0.5,0.5-3.8,2.6);
\drawColormapStackUni{5}{\cmap}{3}{-3.8}{1000}{N_w}

\draw[-stealth,thick] (0.0,-1.85,0) -- (0.5,1.05-3.8,0);
\draw[-stealth,thick] (0.0,-1.85,0.12) -- (0.5,1.05-3.8,1);
\draw[-stealth,thick] (0.0,-1.85,0.5) -- (0.5,1.05-3.8,3);

\begin{scope}[canvas is zy plane at x=0.5]
\draw[black,ultra thick,stealth-] (-0.9,-5.2) -- (-0.9,.8)  node[pos=0.5,sloped,above]{Welch SPOD};
\draw[-stealth, line width=2pt,dash pattern=on 4pt off 1pt] (3,0.5)  to[bend right] (3,0.5-1.7);
\node[rotate=90,font=\footnotesize]  at (2.5,-0.6) {DFT};

\draw[-stealth, line width=2pt,dash pattern=on 4pt off 1pt] (1,0.5)  to[bend right] (1,0.5-1.7);
\node[rotate=90,font=\footnotesize]  at (0.5,-0.6) {DFT};

\draw[-stealth, line width=1.5pt,dash pattern=on 4pt off 1pt] (0,0.5)  to[bend right] (0,0.5-1.7);
\node[rotate=90,font=\footnotesize]  at (-0.5,-0.6) {DFT};

\end{scope}

\begin{scope}[xshift=150,yshift=-10]

\newcommand{\drawColormapStackB}[6]{%
  \pgfmathsetmacro{\steps}{#1}%
  \pgfmathsetmacro{\zOff}{#3}%
  \pgfmathsetmacro{\yOff}{#4}%
    \pgfmathsetmacro{\cstart}{#5}
  \pgfmathsetmacro{\cend}{#6}
  \foreach \step in {0,...,\steps}{%
    \pgfmathsetmacro{\pct}{\cstart + (\cend-\cstart)*\step/\steps}
    \pgfmathsetmacro{\z}{\step/10 + \zOff}%
    \begin{scope}[canvas is xy plane at z=\z]%
      \draw[ellC={\pct of #2}] (0,\yOff) rectangle (1,1+\yOff);%
    \end{scope}%
  }%

}

  \newcommand{\drawColormapStackBoxed}[8]{%
  \pgfmathsetmacro{\steps}{#1}%
  \pgfmathsetmacro{\zOff}{#3}%
  \pgfmathsetmacro{\yOff}{#4}%
  \pgfmathsetmacro{\cstart}{#5}
  \pgfmathsetmacro{\cend}{#6}
    \pgfmathsetmacro{\maxz}{\steps/10+\zOff}%
  \foreach \step in {0,...,\steps}{%
    \pgfmathsetmacro{\pct}{\cstart + (\cend-\cstart)*\step/\steps}
    \pgfmathsetmacro{\z}{\step/10 + \zOff}%
    \begin{scope}[canvas is xy plane at z=\z]%
      \draw[ellC={\pct of #2}] (0,\yOff) rectangle (1,1+\yOff);%
    \end{scope}%
  }%
  
    \draw[#7] (0,0+\yOff,\maxz) -- (0,1+\yOff,\maxz) --(1,1+\yOff,\maxz) -- (1,0+\yOff,\maxz) -- (0,0+\yOff,\maxz) -- (0,0+\yOff,\zOff) -- (0,1+\yOff,\zOff)-- (0,1+\yOff,\maxz);
  \draw[#7] (0,1+\yOff,\zOff) -- (1,1+\yOff,\zOff)-- (1,1+\yOff,\maxz);
  
       \pgfmathsetmacro{\z}{\steps/10 /2 + \zOff}
    \begin{scope}[canvas is xy plane at z=\z]
      \path (0,\yOff) rectangle (1,1+\yOff) node[pos=-0.45,anchor=south west]{$\tilde{\boldsymbol{\mathsf{Q}}}_{#8}$};
    \end{scope}

}

\draw[thick,-stealth] (0,-0.1,0) -- (0,-0.1,3.7) node[pos=1,right]{$t$};

\drawColorGradient{11}{black!70!white}{black!10!white}{0}

\draw[dotted,thick] (0.5,0.5,1.2) -- +(0.0,0.0,1.5);

\begin{scope}[canvas is xy plane at z=1.3]
    \draw[thick,->] (0,0) -- (0,0.5) node[pos=1.2]{$y$};
    \draw[thick,->] (0,0) -- (0.5,0) node[pos=1.4]{$x$};
\end{scope}

\drawColorGradient{5}{black!10!white}{black!0!white}{3}

\drawColormapStackB{11}{\cmap}{0}{-1.7}{0}{700};
\draw[thick,-stealth] (0,-1.8,0) -- (0,-1.8,3.7) node[pos=1,right]{$f$};
\draw[dotted,thick] (0.5,0.5-1.7,1.2) -- +(0.0,0.0,1.5);

\drawColormapStackB{5}{\cmap}{3}{-1.7}{600}{1000}
\drawColormapStackBoxed{5}{\cmap}{0}{-3.8}{0}{300}{black,very thick}{1}
\drawColormapStackBoxed{5}{\cmap}{1}{-3.8}{300}{700}{red,very thick}{2}
\draw[dotted,thick] (0.5,0.5-3.8,2.0) -- +(0.0,0.0,0.5);
\drawColormapStackBoxed{5}{\cmap}{3}{-3.8}{600}{1000}{violet,very thick}{\tilde{N}_w}

\draw[thick, black] (0,-1.9,0) -- ++(0,-0.1,0.0) -- ++(0,0,0.5) -- +(0,0.1,0.0);
\draw[thick, black ,-stealth] (0,-2,0.25) -- (0.5,-3.8+1.05,0);

\draw[thick, red] (0,-1.9,0.55) -- ++(0,-0.1,0.0) -- ++(0,0,0.5) -- +(0,0.1,0.0);
\draw[thick, red ,-stealth] (0,-2,0.75) -- ++(0,-0.05,0) -- (0.5,-3.8+1.05,1);

\draw[thick, red,stealth-stealth] (1,-3.8+1.15,1) -- +(0,0,0.5) node[pos=0.5,above]{$N_f$};

\draw[thick, violet] (0,-1.9,3) -- ++(0,-0.1,0.0) -- ++(0,0,0.5) -- +(0,0.1,0.0);
\draw[thick, violet ,-stealth] (0,-2,3.25) -- ++(0,-0.05,0) -- (0.5,-3.8+1.05,3);

\begin{scope}[canvas is zy plane at x=0.5]
\draw[black,ultra thick,-stealth] (4.2,2) -- (4.2,-4.) node[pos=0.5,sloped,above]{Band-ensemble SPOD};

\draw[-stealth, line width=1.5pt,dash pattern=on 4pt off 1pt] (2,0.5)  to[bend right] (2,0.5-1.7);
\node[rotate=90]  at (2,-0.6) {DFT};

\end{scope}

\end{scope}

\end{tikzpicture}
\caption{Schematic comparison between Welch-based and Band-ensemble SPOD.}\label{fig:schematic}
\end{figure}
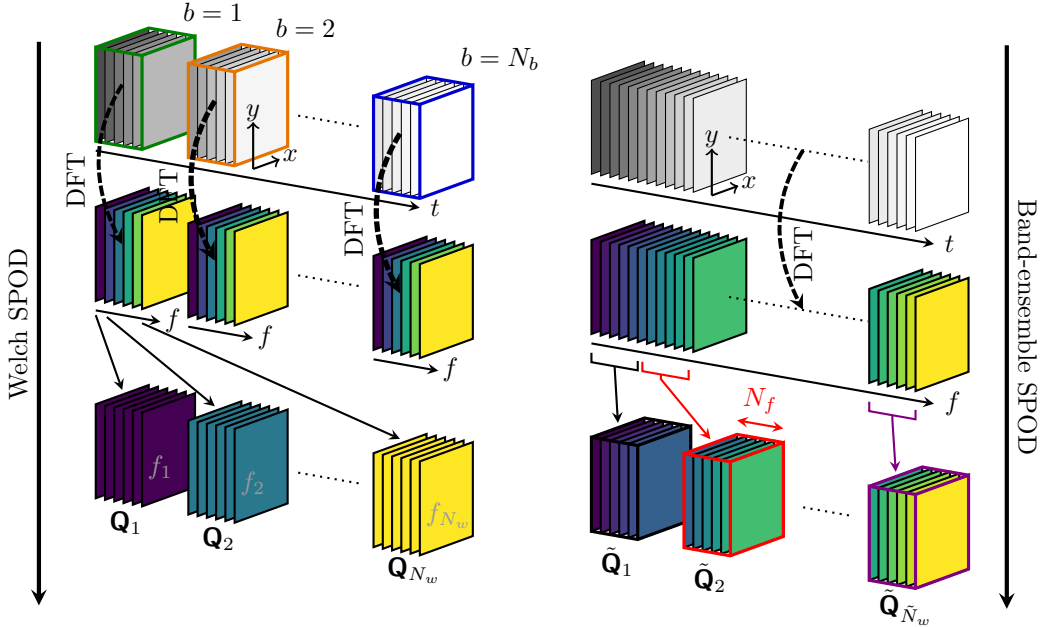

As a final part of the methodology section, we summarize how Welch-based SPOD and the bSPOD approach differ. A schematic of how the data matrices are formed in both cases is shown in Fig.~\ref{fig:schematic}.

The Welch-based algorithm (left) begins by segmenting the time series into blocks and transforming each block into Fourier space via a discrete Fourier transform (black curved arrows labeled DFT). This yields $N_b$ Fourier mode realizations for each frequency. The Fourier modes at a given frequency are then assembled into data matrices $\mathsfbi{Q}_j$, from which the CSD matrix is estimated and the SPOD modes are computed (see Eqs.~\eqref{eq:CSDwelch}-\eqref{eq:SPOD_expansion_method_of_snaps}). The number of frequency bins $N_w$ results from the total signal length $N_t$ and the number of blocks $N_b$. \rev{The CSD estimates and resulting SPOD modes are defined at the respective Fourier-bin frequencies and have an equivalent noise bandwidth of $\Delta f$, assuming rectangular windows.}

bSPOD (right) reverses the order: a single Fourier transform is computed on the full record, yielding Fourier modes on a \rev{finely spaced} frequency grid $(\widetilde{\Delta f})$. Compared to the Welch-based approach without overlap, the \rev{spacing} is $N_b$ times smaller. Block matrices $\tilde{\mathsfbi{Q}}_j$ are then formed from $N_f$ neighboring Fourier modes. The CSD matrix is approximated by summing the contributions of these \rev{narrowly spaced} modes, and the eigenvalue problem is solved to obtain the bSPOD modes (see Eq.~\eqref{eq:CSDb}-\eqref{eq:bSPOD_n_of_snaps2}). Similar to the Welch-based method, the number of \rev{bSPOD} frequency bins $\tilde{N}_w$ results from the total time length $N_t$ and the filter lengths $N_f$. \rev{The effective bandwidth of the CSD estimate and the corresponding SPOD modes spanned by the ensemble of $N_f$ consecutive Fourier modes is $N_f \widetilde{\Delta f}$.}

If the number of consecutive Fourier modes $N_f$ used in bSPOD is equal to the number of blocks $N_b$ in Welch-based SPOD, the resulting CSD matrices have the same size and represent the cross-spectral density \rev{with} the same \rev{effective} frequency \rev{bandwidth,} $\Delta f = N_f\,\widetilde{\Delta f}$. Under the usual SPOD assumption that the flow dynamics \rev{associated with the contributing spectral content remain sufficiently similar}, the two approximations should yield the same results. 

\rev{The key distinction is that Welch-based SPOD relies on temporal segmentation to estimate the CSD matrix, whereas bSPOD estimates modes from contributions of consecutive Fourier coefficients defined on a finer frequency grid obtained from a single Fourier transform of the full-length record. This difference gives rise to several theoretical and practical implications.}
\rev{First, the different approaches used to estimate the CSD lead to a different interpretation of the SPOD expansion coefficients. In Welch-SPOD, the coefficients quantify the contribution of individual time blocks to a mode. In bSPOD, the coefficients instead quantify the contribution of neighboring Fourier coefficients within a frequency band. This frequency-resolved representation enables the frequency-attribution procedure introduced above and provides additional information about the spectral distribution of modal energy within a band.}

\rev{Second, Welch-based SPOD employs block segmentation and assumes statistical independence between blocks. In contrast, bSPOD does not introduce this assumption and instead relies on a Fourier transform of the complete time signal. Consequently, bSPOD relies on a single long observation period over which the statistics remain sufficiently stationary and is not directly suited for combining multiple shorter records of the same process.}
\rev{In the limit of infinite observation time, Fourier coefficients at distinct frequencies are theoretically uncorrelated, thereby providing independent contributions to the approximation of the CSD matrix. While finite record lengths introduce correlations between neighboring Fourier coefficients through spectral leakage, these effects decrease with increasing record length. Consequently, neighboring Fourier coefficients provide approximately independent realizations for sufficiently long time series.}

\rev{For $N_f=N_b$, both methods define CSD matrices associated with the same effective bandwidth $\Delta f$. In a direct comparison, however, it should be noted that Welch-SPOD associates the mode estimates with the Fourier-bin frequencies, Eq.~\ref{eq:fourier_bins}, whereas bSPOD defines each estimate over an interval spanned by $N_f$ consecutive Fourier modes. Without overlap and starting the ensembles from the first Fourier mode, this results in the bSPOD frequency grid being shifted by approximately $\Delta f/2$ relative to the Welch-SPOD grid, for which the effective bandwidth is represented only implicitly. In practice, however, this distinction is of little importance, as the objective of both methods is to estimate the modal power spectral density as a function of frequency.}

\rev{Moreover, the bSPOD algorithm design entails a slightly different computational cost profile.}
A single Fourier transform of length $N_t$ is more expensive than $N_b$ Fourier transforms of length $N_t/N_b$. However, bSPOD avoids overhead costs associated with segmentation, tapering, and looping over blocks. Additional computational cost may arise because bSPOD relies on the method of snapshots, which is more expensive when the number of spatial degrees of freedom is small compared to the frequency-window length $N_f$, and because of the additional operations required for frequency attribution. Nevertheless, the overall computational cost remains comparable to that of Welch-based SPOD, since the eigenvalue problem, typically the most expensive step, is of identical size for comparable configurations ($N_b=N_f$). In return, bSPOD offers clear advantages: performing the Fourier transform over the full time series reduces spectral leakage, and the frequency-attribution step enables improved frequency estimations for tonal components as demonstrated in Section~\ref{sec:results}. Lastly, the \rev{effective bSPOD} frequency \rev{resolution} can be varied without recomputing the Fourier transform, making the approach suitable for adaptive frequency resolution strategies~\cite{yeung2024}.

\section{Databases}\label{sec:data}
\subsection{Artificial \rev{broadband-tonal} signal}\label{sec:artdata}
To assess the performance of bSPOD, we construct an artificial space--time signal that is statistically stationary and contains both broadband and tonal features. The goal is to mimic typical characteristics of turbulent flows while retaining full control over the spectral content. The signal is defined on a periodic one-dimensional domain of length $L_x$, sampled at $N_x$ equispaced points. Time is discretized into $N_t$ samples with time step $\Delta t$. The corresponding grids are
\begin{equation}
x_l = l\,\Delta x, \quad l=0,\dots,N_x-1, 
\qquad 
t_n = n\,\Delta t, \quad n=0,\dots,N_t-1,
\end{equation}
with $\Delta x=L_x/N_x$ and $\Delta t=T/N_t$. The broadband part of the signal is obtained by superimposing two independent signals $y_m$, $m=1,2$, each representing a convecting wavepacket with random phase and amplitude. The target temporal spectrum for both signals is chosen as the response of a single oscillator with natural frequency 
$f_{0}$ and damping $\zeta$:
\begin{equation}
S(\omega) = 
\frac{4 \zeta^2 \omega_{0}^4}{\big[\omega_{0}^2-\omega^2\big]^2
+ \big[2\zeta \omega_{0}\omega\big]^2}
\end{equation}
where $\omega=2\pi f$ denotes the angular frequency.

The signal is constructed in frequency space and then inverse transformed to
the physical domain. At each discrete frequency $\omega_j = j\,2\pi/T$ we draw
independent complex Gaussian coefficients,
\begin{equation}
A(\omega_j) = \sqrt{S(\omega_j)\,\Delta f}\; z_{j},
\qquad 
z_{j} = \frac{\big(z_{\mathrm{re}}+ i\,z_{\mathrm{im}}\big)}{\sqrt{2}},
\end{equation}
where $z_{\mathrm{re}},z_{\mathrm{im}}\sim \mathcal{N}(0,1)$ are independent 
standard normal draws and $\Delta f=1/T$ is the frequency bin width. This 
construction ensures $\mathbb{E}[|A(\omega_j)|^2] = S(\omega_j)\,\Delta f$, 
so that the expected power spectrum matches the target, while the random phase 
renders the realization statistically stationary. The scaling with $\Delta f$ ensures that the variance of the signal equals the integral of the prescribed power spectral density. Consequently, apart from the random realization, both broadband signals share the same spectral distribution, while differences arise primarily from their spatial structure, which is considered next. 

To introduce spatial coherence, each frequency component is assigned a
wavenumber through the linear dispersion relation
\begin{equation}
k_m(\omega_j) = \frac{\omega_j}{c_m} + k_{0,m},\label{eq:dispersion_relation}
\end{equation}
modeling convection at speed $c_m$ with offset $k_{0,m}$, which are chosen differently for the two broadband components. To avoid introducing a preferred spatial origin, the two broadband branches are displaced by a random shift $x_{0,m} \sim \mathcal{U}(0,L_x)$, where $\mathcal{U}$ denotes the uniform distribution. This ensures that the ensemble is statistically homogeneous in space.
The resulting signals read
\begin{equation}
y_m(x_l,t_n) = \sum_{j=1}^{ N_t/2 }
A_m(\omega_j)\,
\exp\!\Big(i\big[k_m(\omega_j)(x_l-x_{0,m}) + \omega_j t_n\big]\Big).
\end{equation}
To complement the broadband signal, we superimpose five discrete tones:
\begin{equation}
o_r(x_l,t_n) = B_r 
\exp\!\Big(
i\big[k_r x_l/L_x + \omega_r t_n + \phi_r\big]
\Big), \quad r=1,\dots, 5
\end{equation}
with amplitude $B_r$, frequency $\omega_r$, spatial wavenumber $k_r$, and random 
phase $\phi_r \sim \mathcal{U}(0,2\pi)$. These time series generate sharp spectral peaks adding tonal components to the broadband signals.
The complete artificial signal is obtained by superimposing the broadband and the tonal components
\begin{equation}
y(x_l,t_n) = \sum_{m=1}^{2} g_m\, y_m(x_l,t_n) + \sum_{r=1}^{5} o_r(x_l,t_n),
\end{equation}
with prescribed scalar weights $g_m$ for the two broadband signals to separate them into a leading and a sub-leading mode.

In the results shown, we set $N_x=250$, $L_x=10$, $N_t=100{,}000$, 
$\Delta t=10^{-3}$.  
The broadband field contains two branches with parameters 
$f_0=100$, 
$\zeta=0.2$, $(g_m,c_m,k_{0,m}) \in \{(1.0,200,0.5),\,(0.5,150,1.5)\}$.  
Five tones are added with $(B_r,f_r,k_r) \in$ 
$
\{(5,23,3),(3,25,8),$ $(6,30,5),
(5,75,3),(5,125,6)\}
$.
 
\subsection{Broadband-tonal cavity flow}\label{sec:cavityflow}
To demonstrate the methodology on experimental data, bSPOD is applied to a high-speed mono-PIV data set of a turbulent open cavity flow~\citep{zhang2020}. 
This data set has previously been used to benchmark spectral estimation 
techniques, for example the adaptive SPOD algorithm based on multitaper 
estimation for improved tone identification~\citep{yeung2024}, and by \citet{zhang2020}, who estimate spectral POD modes of non-time-resolved PIV snapshots by making use of high-speed pressure measurements.

\begin{figure}
    \centering
    \includegraphics[width=0.49\linewidth]{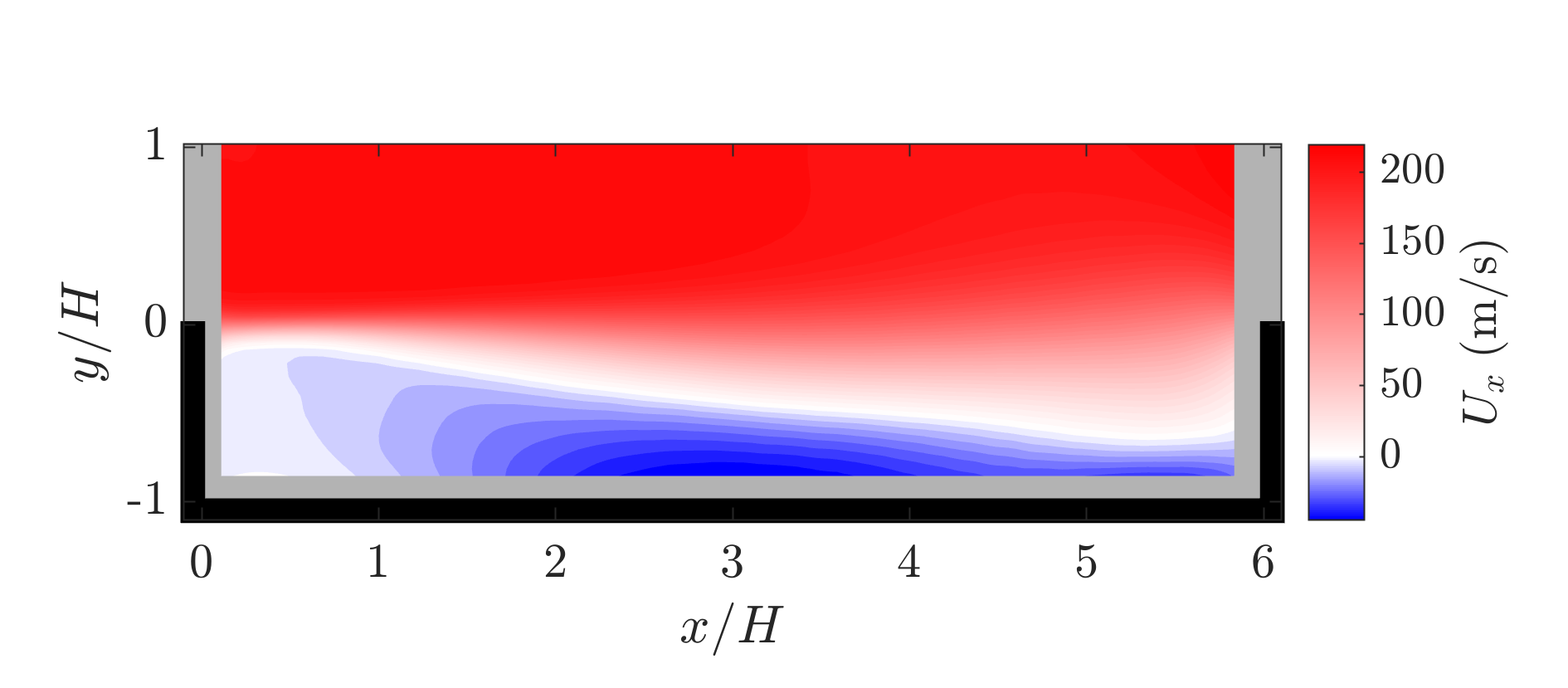}
    \includegraphics[width=0.49\linewidth]{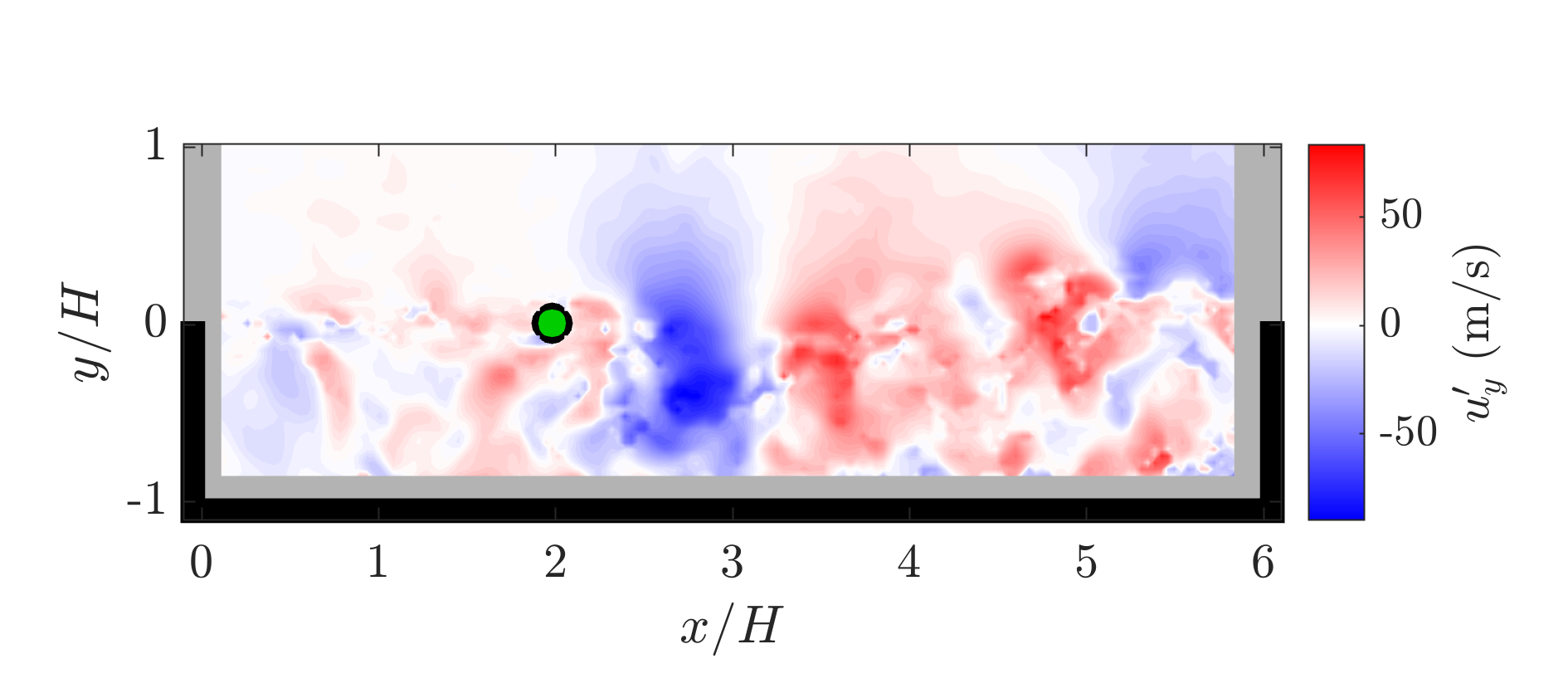}
    \includegraphics[width=0.49\linewidth]{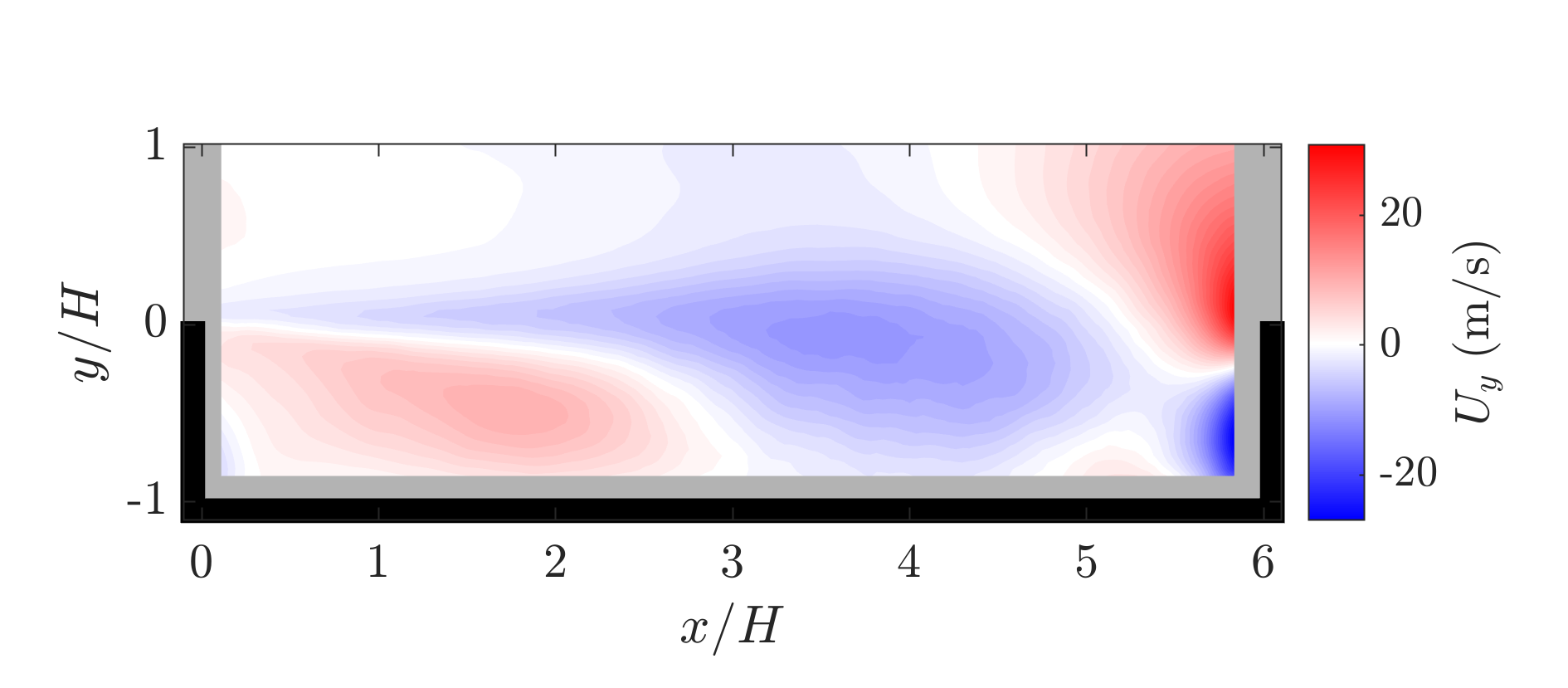}
    \includegraphics[width=0.49\linewidth]{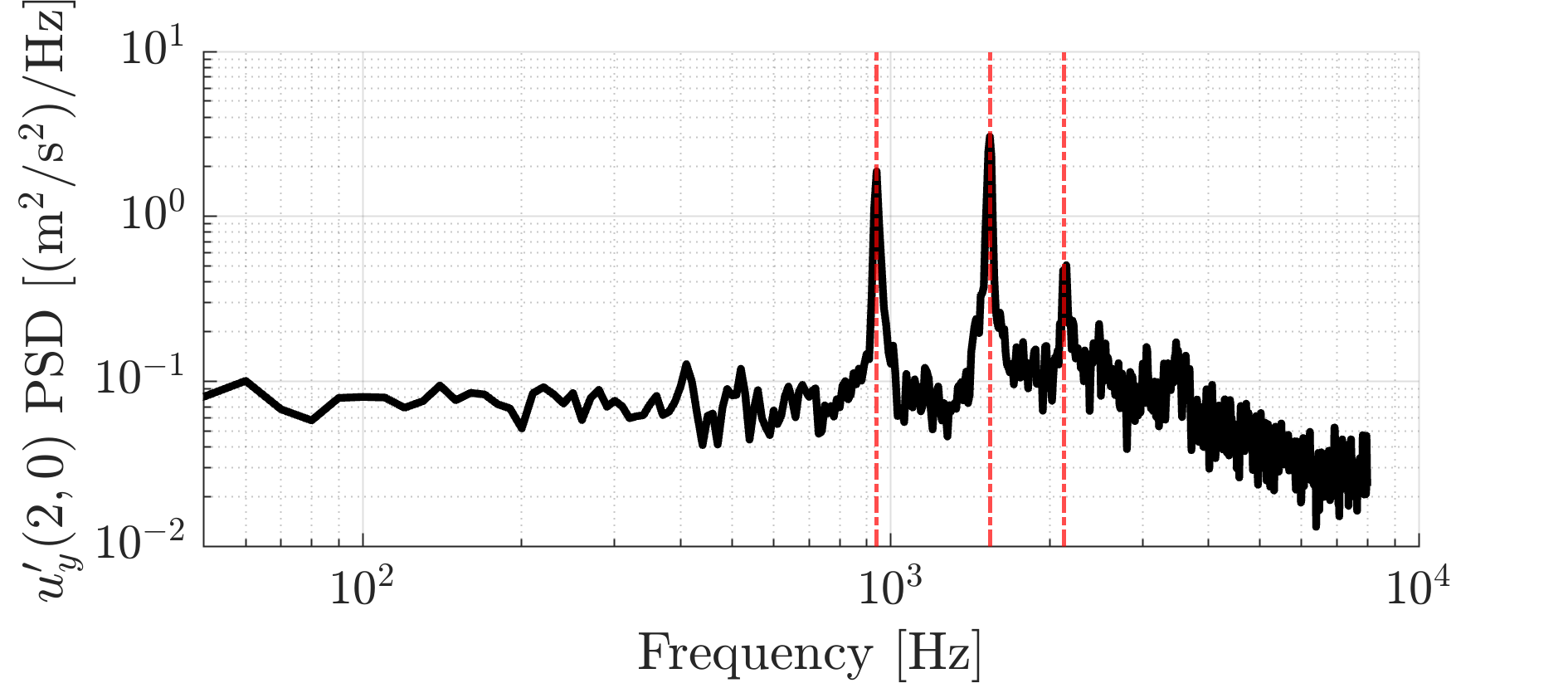}
\caption{Mach 0.6 cavity-flow mono-PIV dataset of~\citet{zhang2020}. Left: time-averaged axial (top) and vertical (bottom) velocity field; thick black lines mark the cavity boundaries. Right: snapshot of the fluctuating (mean-subtracted) vertical velocity (top) and the PSD (bottom) of the fluctuating vertical velocity at the green dot. Red vertical lines indicate the three Rossiter modes (940, 1540, and 2130 Hz).}
    \label{fig:cavity_fields}
\end{figure}

The cavity has a depth of $H=26.5$~mm, an axial extent of $6H$, and an 
out-of-plane width of $3.85H$. The PIV measurements provide the \rev{center-plane} axial and vertical velocity components on an equidistant grid of $N_x=154$ by $N_y=54$ points, with a total of $N_t=16{,}000$ snapshots sampled at $\Delta t=6.25\times 10^{-5}$s. 
Figure~\ref{fig:cavity_fields} shows the time-averaged axial and vertical velocity fields, a snapshot of the fluctuating vertical velocity component $u_y^\prime$, and the spectral power density of the fluctuating vertical velocity component $u_y^\prime (2,0)$. The location at which this component is extracted is highlighted with a marker in the $u_y^\prime$ plot on the top right. The PSD is computed using Welch’s method with a window length of 1600 samples and 50\% overlap. Three distinct peaks are observed at 940, 1540, and 2130~Hz, corresponding to the second, third, and fourth Rossiter modes.

\section{Results}\label{sec:results}

\subsection{Characteristics of bSPOD}
\begin{figure}
    \centering
    \includegraphics[width=0.95\linewidth]{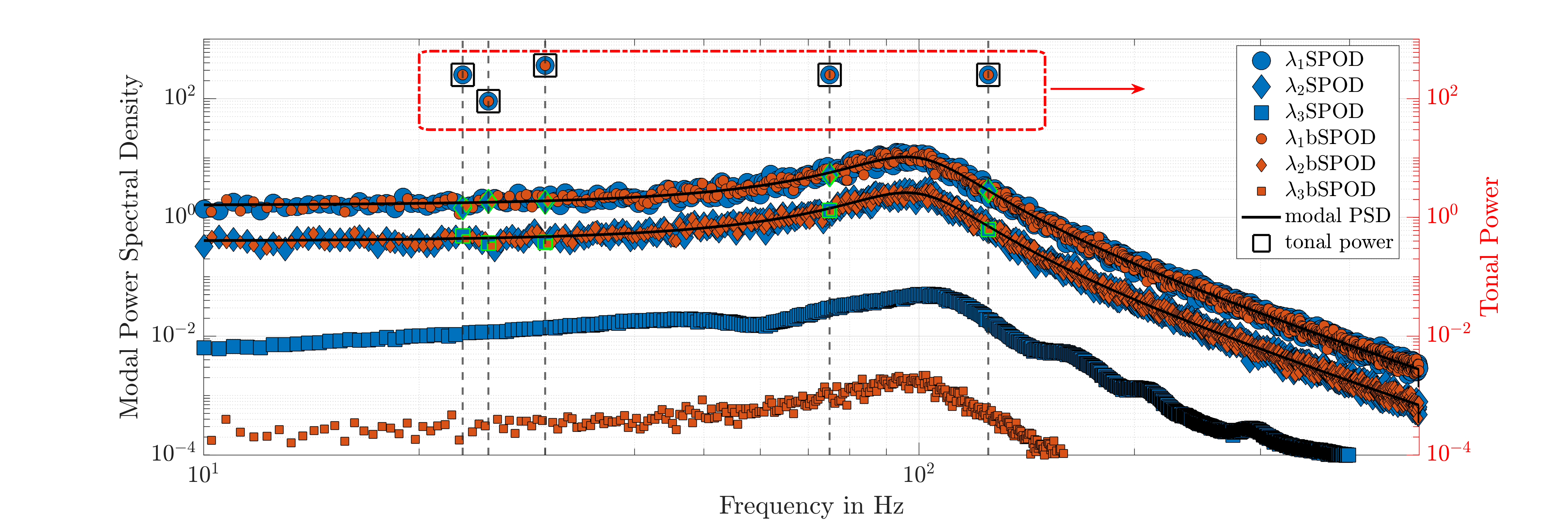}
    \includegraphics[width=0.95\linewidth]{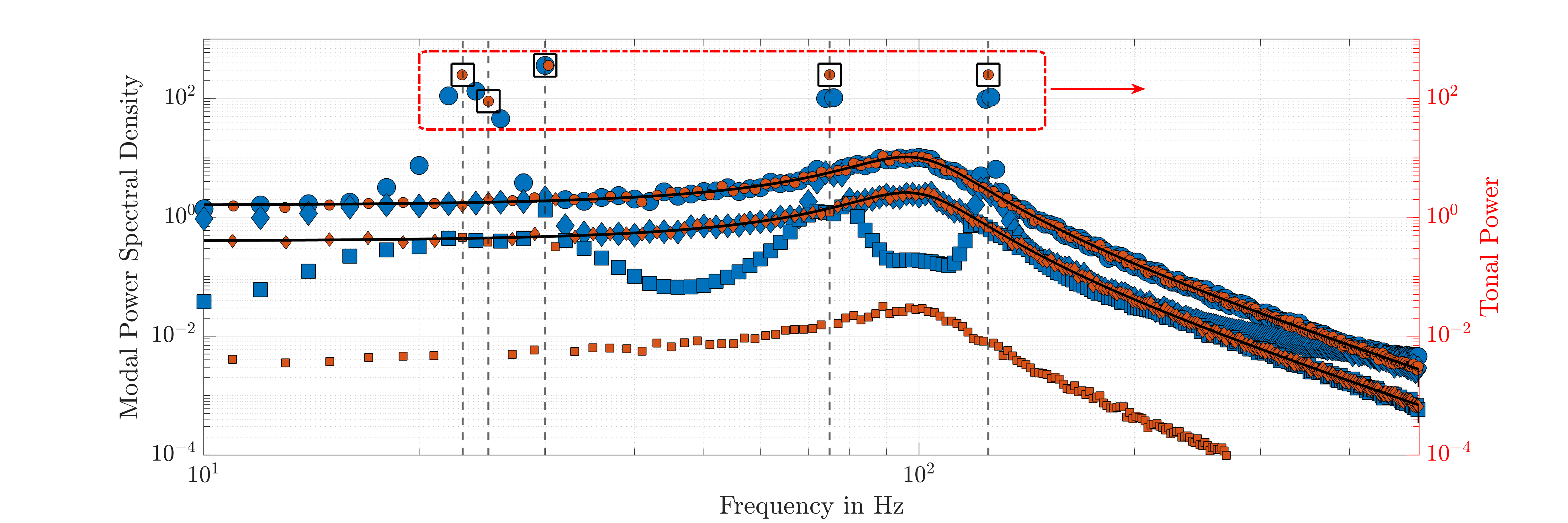}
    \includegraphics[width=0.95\linewidth]{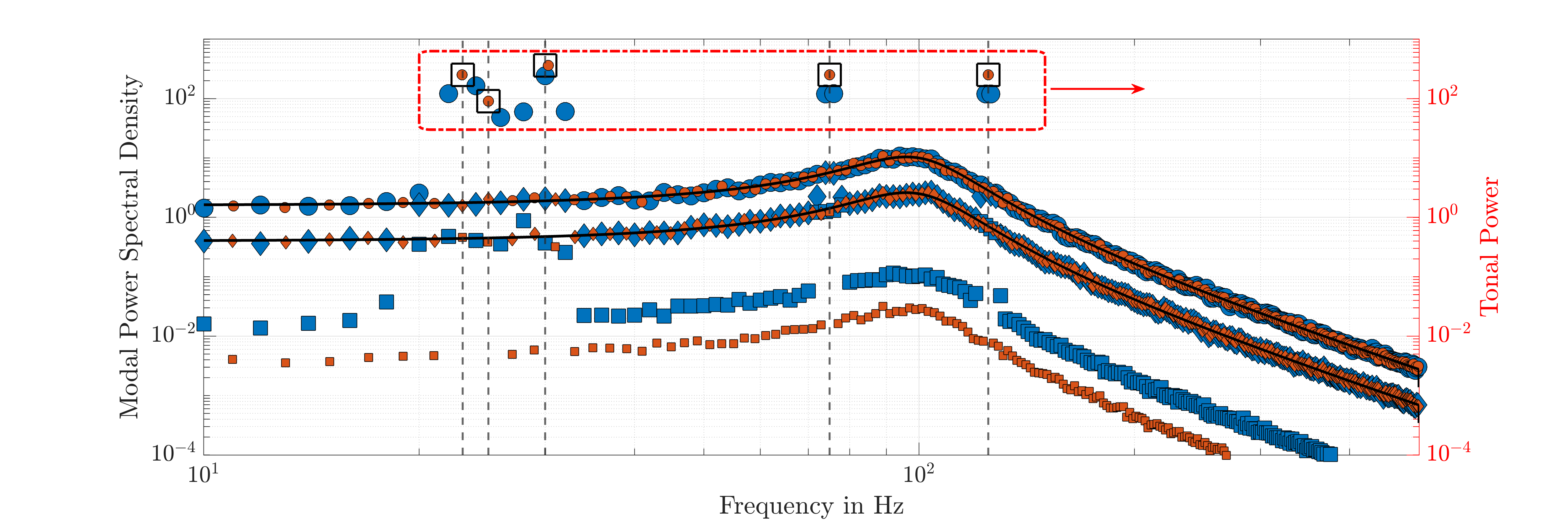}
    \caption{Modal power spectral densities and tonal power for the artificial data set. Welch-based SPOD and bSPOD are compared against the theoretical values for three configurations: $\Delta f=0.5,\mathrm{Hz}$ with $N_b=N_f=50$ (top), $\Delta f=2\,\mathrm{Hz}$ with $N_b=N_f=200$ (middle), and $\Delta f=2\,\mathrm{Hz}$ with $N_b=N_f=200$ using a Hann taper for Welch-based SPOD (bottom). Blue and orange markers denote the leading three eigenvalues from Welch-based SPOD and bSPOD, respectively.  Markers enclosed by the red box represent tonal power (right axis), whereas all other markers and lines indicate power spectral densities (left axis). Black lines show the analytic broadband power spectral density ($g_m^2 S(\omega)L_x$), and black squares indicate the true tonal power ($B_r^2L_x$), see Section~\ref{sec:artdata}. Dashed vertical lines mark the tonal frequencies.}
    \label{fig:artificial1}
\end{figure}

To highlight the features of bSPOD, we first consider the artificial broadband–tonal data set. The top panel in Fig.~\ref{fig:artificial1} shows the eigenvalue spectra of the first three modes from both Welch-based SPOD and bSPOD, together with the analytic ground-truth values known from the construction of the signal. For the broadband components $y_m$, the spatially integrated power spectral density is shown; for the tonal components $o_r$, the integrated tonal power is shown, since their energy is concentrated at single frequencies. SPOD and bSPOD results are shown for $N_b=N_f=\,$50, which yields a fine frequency resolution of $\Delta f=N_f\widetilde{\Delta f}=\,$0.5~Hz. Both methods are initially applied without overlap or windowing to compare the core algorithms without additional features. Within each frequency band, the bSPOD modes are assigned to frequencies according to Eq.~\eqref{eq:weighted_freqs} with weight exponent $\ell=1$. Since the tonal component's spectral densities correspond to Dirac impulses, we scale the associated peak eigenvalues from both SPOD variants by the \rev{frequency spacing} $\Delta f$ to enable direct comparison with the tones’ total power. The corresponding eigenvalues representing tonal power are enclosed by a red box in Fig.~\ref{fig:artificial1}.

Both methods accurately capture the frequency and power of the tones. 
The spectra of the two broadband components are also represented well, although a high estimation variance is observed, visible in the noisy distribution of the eigenvalue estimates (markers) around the expected ground-truth values (black solid lines). The high variance results from the small number of blocks and the short band-ensemble filter width $N_b = N_f$, reflecting the expected trade-off between frequency resolution and statistical convergence.

The leading and first subleading broadband modes are reproduced by the respective leading and subleading SPOD (bSPOD) eigenvalues. In the bins containing tones, the dominant SPOD (bSPOD) mode represents the tone, shifting the mode ranking by one: the leading broadband component appears as the first subleading mode, the subleading brodband component by the second subleading mode, and so on. These shifted SPOD and bSPOD eigenvalues are highlighted in the top panel of Fig.~\ref{fig:artificial1} with green marker edges.

Overall, the differences between the shown SPOD and bSPOD results are small, except that bSPOD shows less spectral leakage. 
This can be observed from the magnitude of the second subleading mode (square markers), which is purely a result of leakage and is about one order of magnitude lower for bSPOD than for SPOD. This reduction follows from using the full time series in the Fourier transform, as opposed to shorter windows in Welch-based SPOD with time segmentation.

To reduce the estimation variance, the number of blocks and the band-ensemble filter length are increased by a factor of four, which decreases the frequency resolution of both methods by the same factor. The corresponding eigenvalue spectra are shown in the middle plot of Fig.~\ref{fig:artificial1}. For bSPOD the results are plotted \rev{again for $\ell=1$}, Eq.~\eqref{eq:weight}. As in the upper figure with higher frequency resolution, bSPOD reproduces both tonal and broadband components accurately, now with significantly reduced variance for the broadband signals. 
By contrast, the Welch-based SPOD eigenvalues deviate from the analytic reference: the frequency \rev{resolution is insufficient} to resolve the tones, and the short windows lead to severe leakage effects. 
This affects not only the regions near the tonal peaks but also the subleading modes at higher frequencies, where leakage obscures the separation between physical modes and spurious ones.

In Welch-based SPOD, spectral leakage is commonly mitigated by applying a taper to each time segment. The blue markers in the bottom panel of Fig.~\ref{fig:artificial1} show the corresponding eigenvalues obtained when a Hann window is used as the taper. The bSPOD eigenvalues are unchanged relative to the middle panel and are replotted again for direct comparison. Tapering substantially reduces spectral leakage for frequencies away from the tonal ones; however, the spurious secondary subleading-mode branch induced by leakage remains more pronounced for Welch-based SPOD than for bSPOD. 
Furthermore, a side effect of tapering in Welch-based SPOD is increased leakage in the immediate vicinity of the tones. This follows because windowing corresponds to convolution with the window’s frequency response: compared to the rectangular window, the Hann window has a broader main lobe, which increases spectral smearing near the tonal peaks~\citep{harris1978}.

In bSPOD, leakage is naturally reduced by computing the Fourier transform over the full time series, without the need for tapering. Applying a taper within bSPOD would further reduce the amplitude of the spurious mode branch.
Overall, these results demonstrate that bSPOD yields improved frequency selectivity and reduced spectral leakage relative to the Welch-based formulation, even when tapering is applied.

\subsection{Application to experimental data}
After applying the proposed method to an artificial test data set, we turn to the more relevant case and apply Welch-based SPOD and bSPOD to the mean-subtracted cavity flow velocity snapshots, $u_x'$ and $u_y'$. For a fair comparison, the methods are again configured with $N_f = N_b$; that is, the number of blocks in Welch-based SPOD equals the number of consecutive Fourier modes considered in bSPOD. No overlap is applied, ensuring that both methodologies produce the same number of mode estimates \rev{with the same effective frequency spacing $\Delta f$}. For Welch-based SPOD, a Hann window is used to reduce spectral leakage, while in bSPOD the frequency attribution is based on $\ell=\infty$, see Eq.~\eqref{eq:weight}.

\begin{figure}
    \centering
    \includegraphics[width=0.95\linewidth]{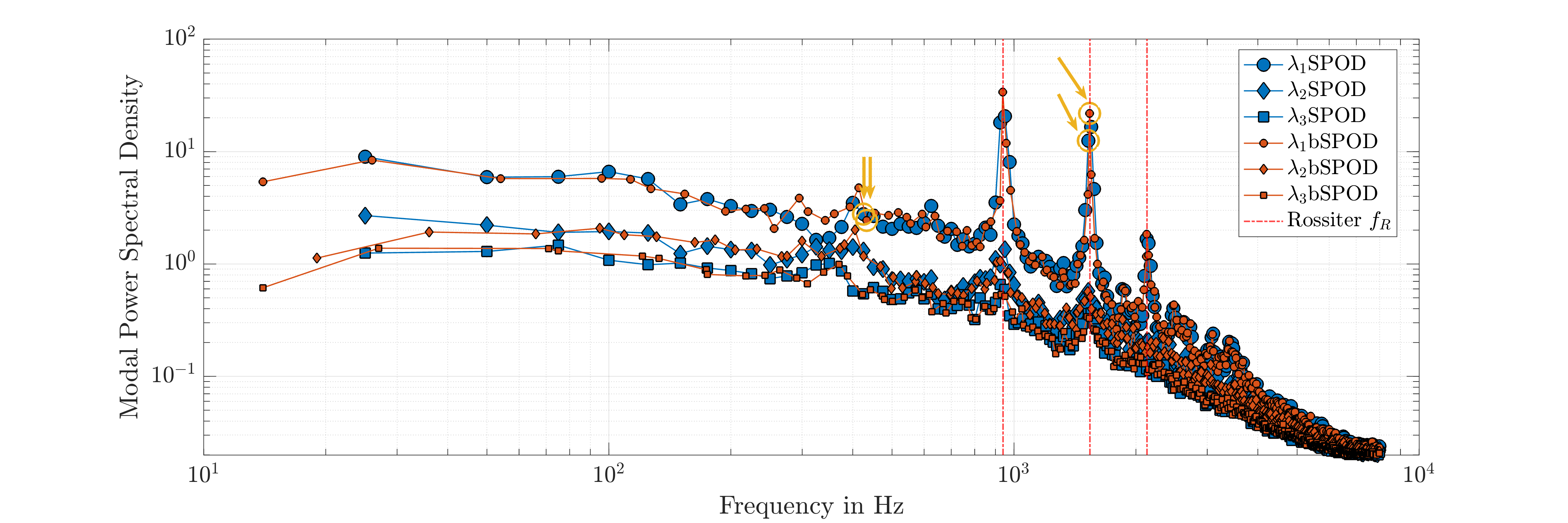}
     \includegraphics[width=0.95\linewidth]{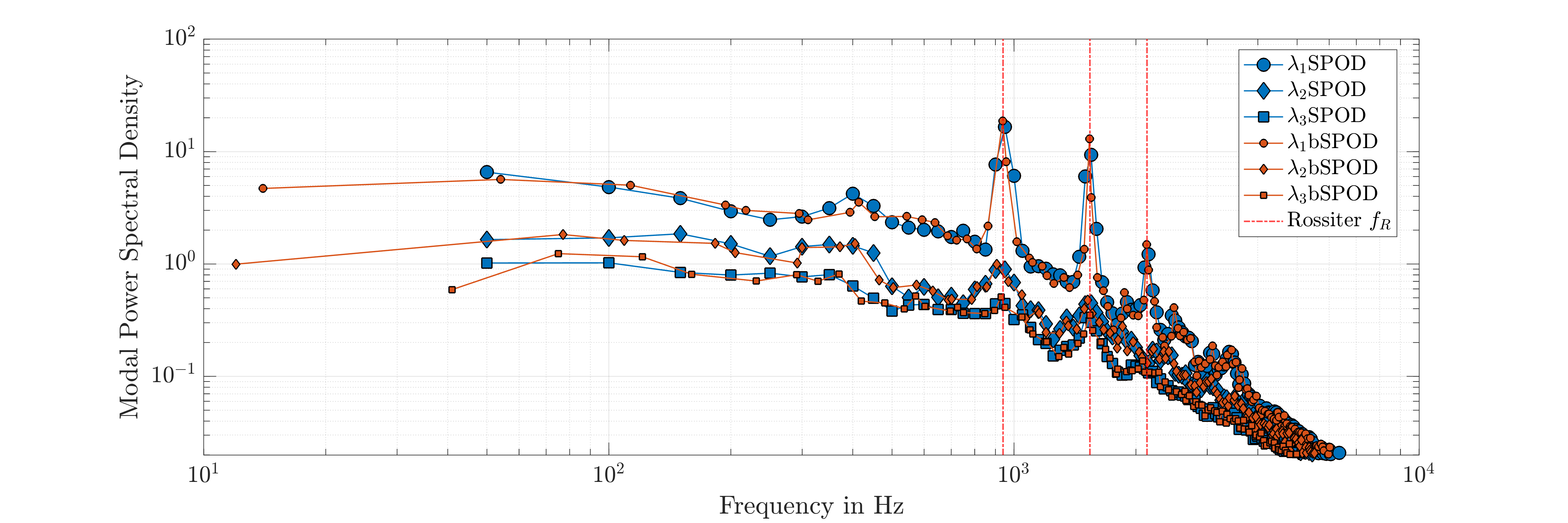}
    \includegraphics[width=0.95\linewidth]{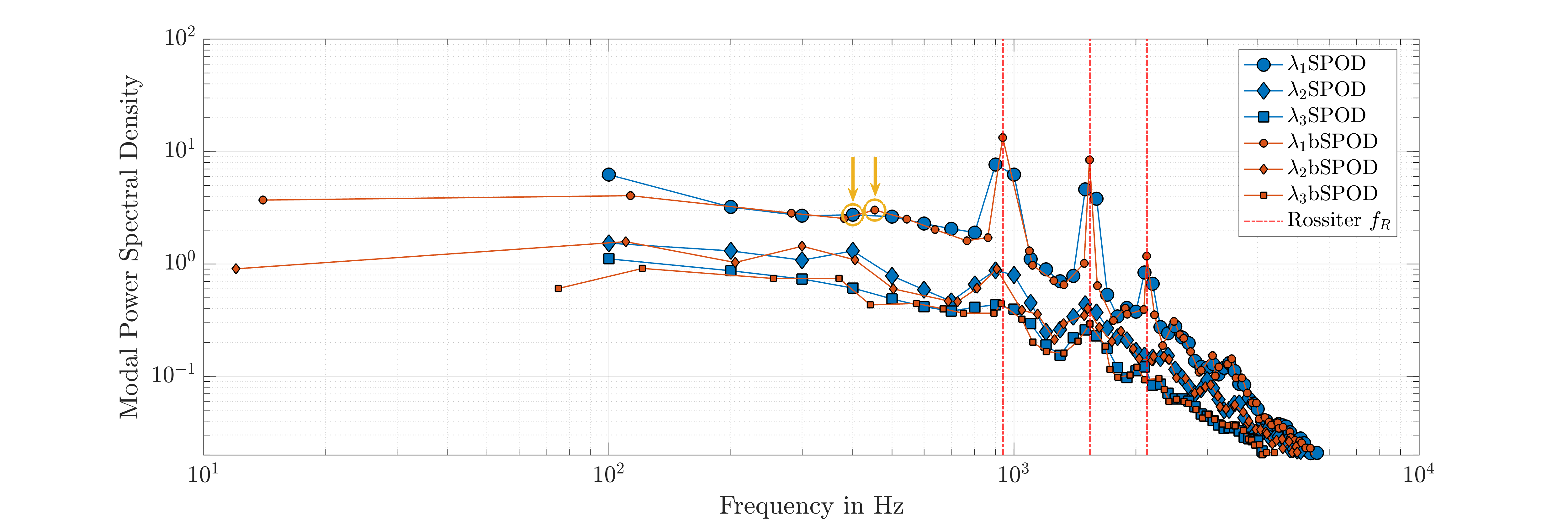}
    \caption{Leading three SPOD eigenvalues versus frequency for the Welch-based (blue) and band-ensemble (orange) formulations with $N_f=N_b$ and three frequency resolutions: $\Delta f=25~\mathrm{Hz}$ (top), $50~\mathrm{Hz}$ (middle), and $100~\mathrm{Hz}$ (bottom). Vertical red lines denote the Rossiter-mode frequencies identified from the single-point PSD (see Fig.~\ref{fig:cavity_fields}). Yellow arrows and circles mark the eigenvalues whose corresponding mode shapes are shown in Section~\ref{sec:mode_convergence}.}
    \label{fig:cavity1}
\end{figure}

The top panel of Fig.~\ref{fig:cavity1} shows the power spectral density of the three most energetic modes obtained with both methods for a frequency resolution of $\Delta f=\,$25~Hz ($N_b=N_f=25$). The dashed vertical lines highlight the frequencies of the three Rossiter modes, obtained from a point-wise PSD as described in \S~\ref{sec:cavityflow}. At this high frequency resolution, the three tonal Rossiter peaks are captured well by both approaches. However, the high frequency resolution comes at the cost of large estimation variance, visible as elevated noise levels in regions away from the tones where the leading mode represents broadband dynamics. The observed noise level indicates that the spectra are not well converged and more modes are required which, however, would reduce the frequency resolution. 
This trade-off is illustrated in the middle and bottom panels of Fig.~\ref{fig:cavity1}, which show the same spectra for $N_b=N_f=50$ ($\Delta f=50$~Hz) and $N_b=N_f=100$ ($\Delta f=100$~Hz), respectively. 
Although both methods experience the same decrease in frequency resolution, bSPOD maintains accurate tonal frequency estimates enabled by the data-driven frequency attribution.
In contrast, such in-band frequency information is not available for Welch-SPOD. Furthermore, the Welch approach remains prone to spectral leakage: energy from neighboring tones contaminates the spectral estimate, resulting in a biased representation of the tonal components. These results highlight the advantages of bSPOD for analyzing broadband--tonal flows. 
By reducing spectral leakage and exploiting the information contained in the expansion coefficients, bSPOD achieves sharper spectral localization and a clearer separation of tonal and broadband contributions than the Welch-based SPOD formulation.

\subsection{Mode convergence}\label{sec:mode_convergence}
\begin{figure}
    \centering
     \includegraphics[width=0.49\linewidth]{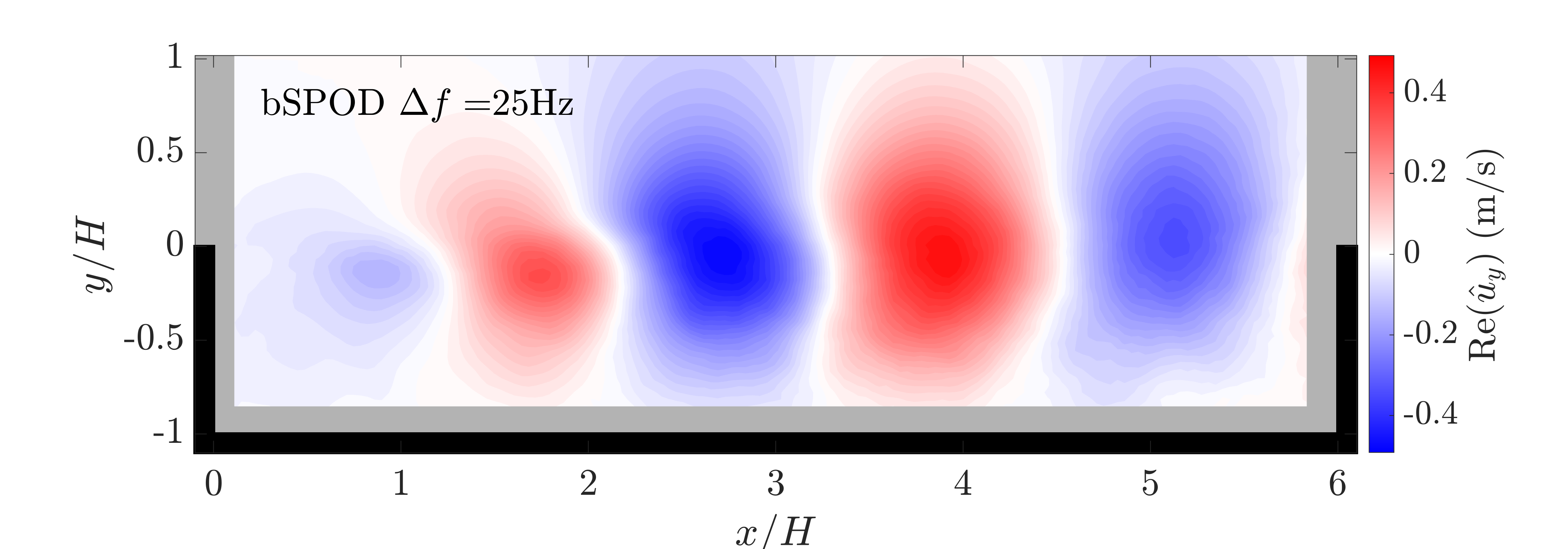}
    \includegraphics[width=0.49\linewidth]{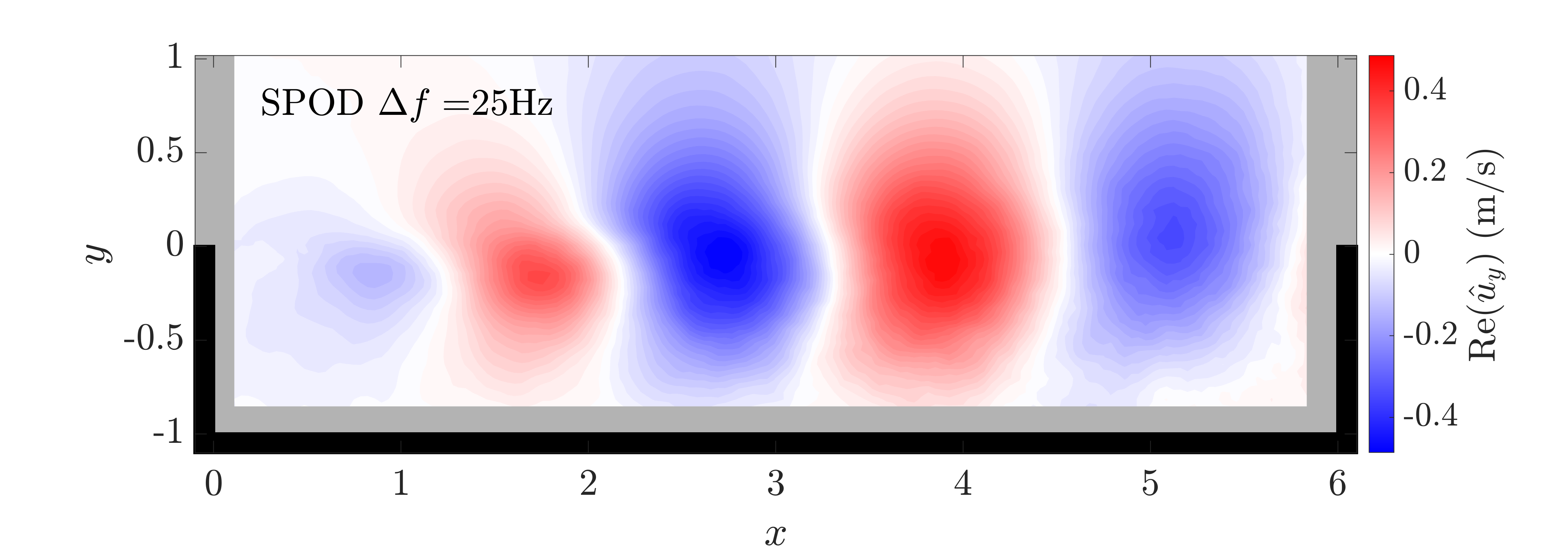}
    \caption{Real part of the vertical velocity component of the Rossiter mode at $1540\,$Hz obtained with band-ensemble SPOD (left) and Welch-based SPOD (right), both using \rev{an effective} frequency resolution of $\Delta f=25\,$Hz. The corresponding eigenvalues are marked by yellow crosses in the top panel of Fig.~\ref{fig:cavity1}.}
    \label{fig:cavity2}
\end{figure}

\begin{figure}
    \centering
    \includegraphics[width=0.49\linewidth]{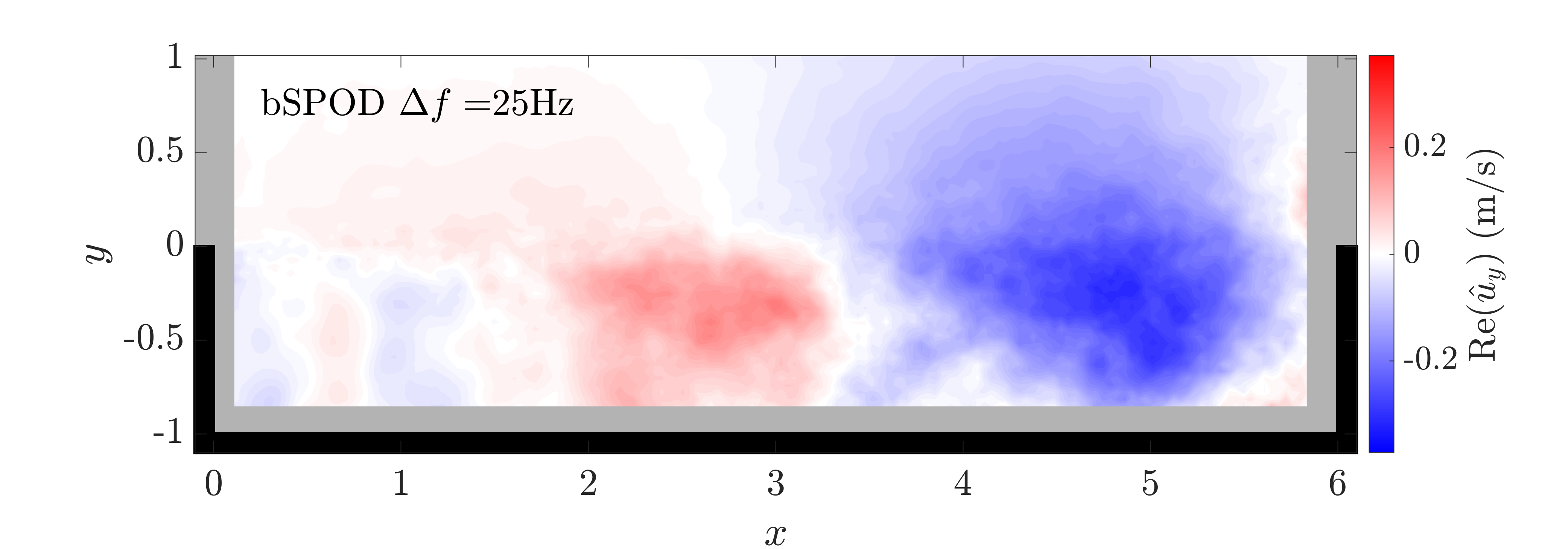}
     \includegraphics[width=0.49\linewidth]{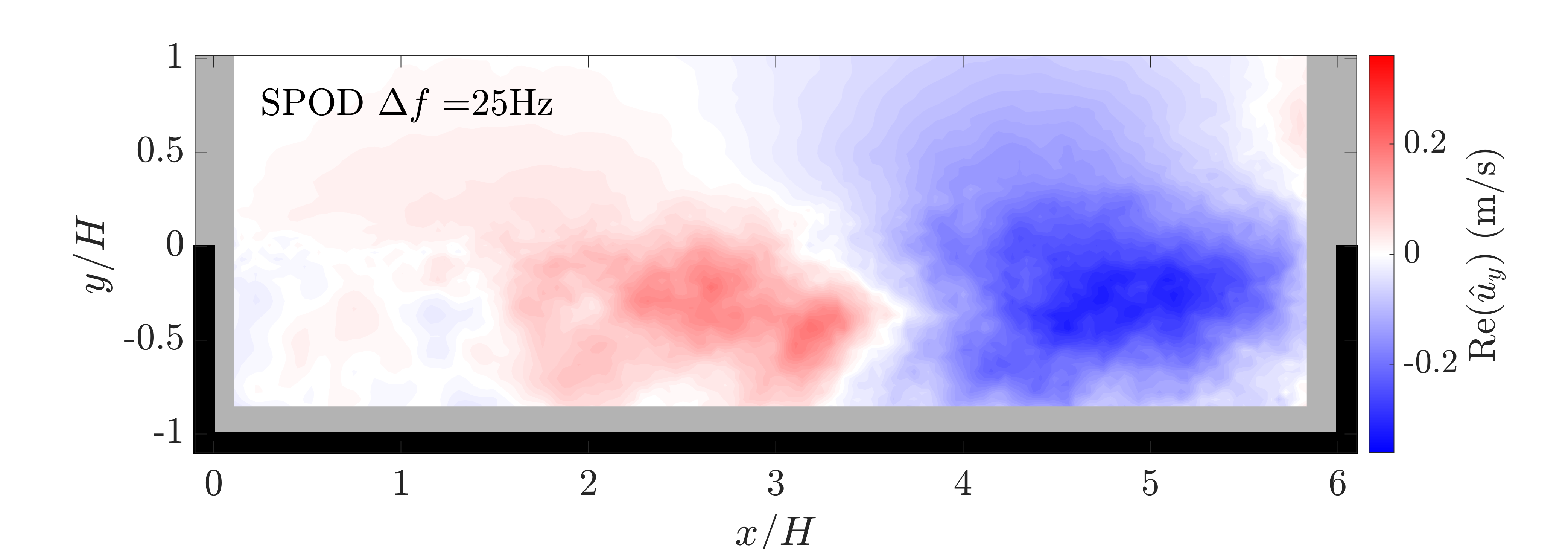}
     \includegraphics[width=0.49\linewidth]{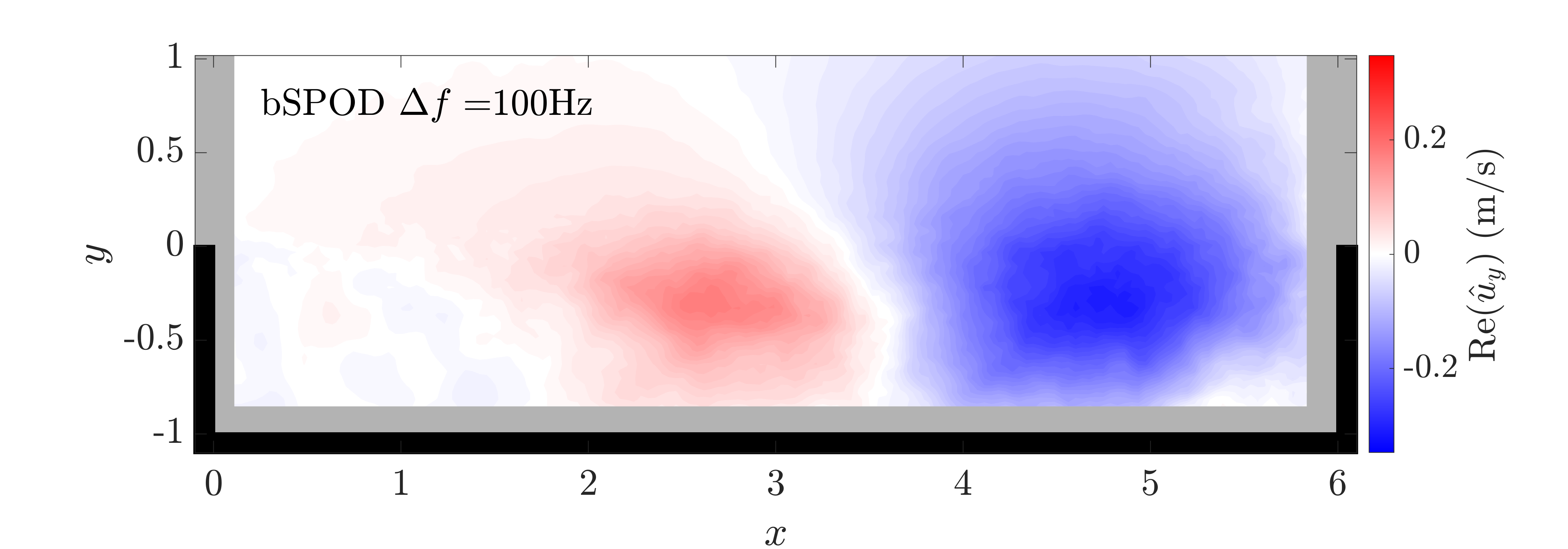}
    \includegraphics[width=0.49\linewidth]{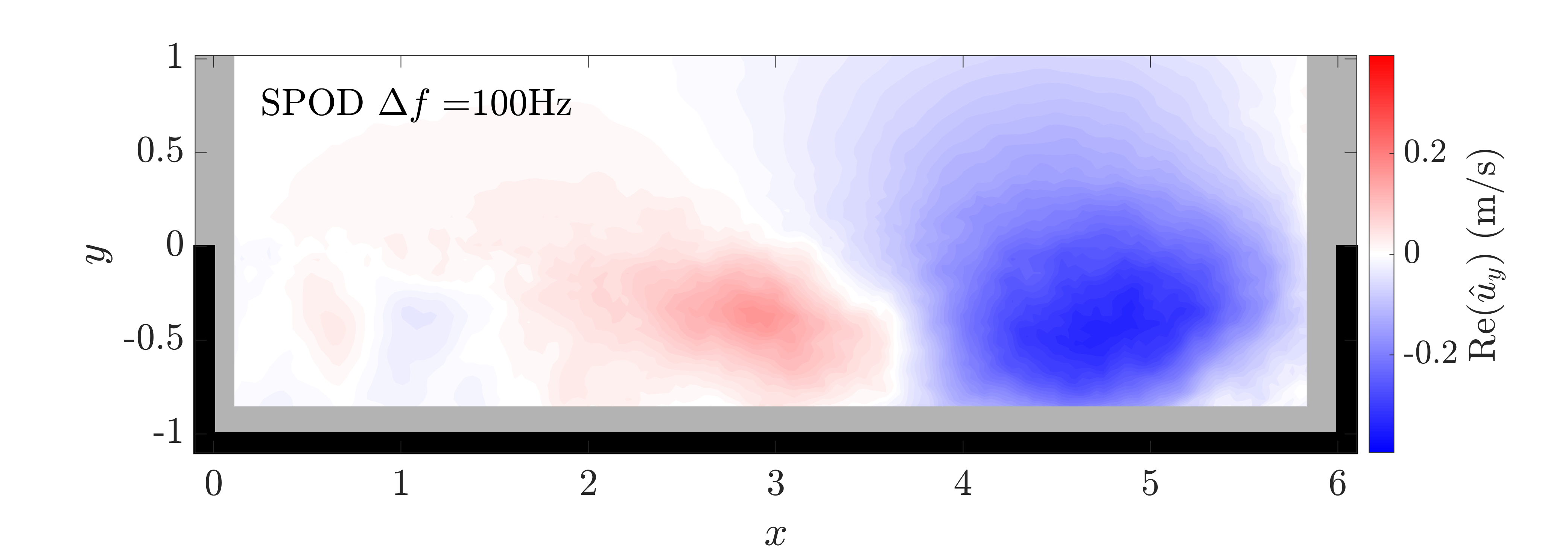}
    \caption{Real part of the vertical velocity component of the leading broadband mode computed at two \rev{effective} frequency resolutions: $425$–$450\,$Hz (top) and $400$–$500\,$Hz (bottom). Results are shown for band-ensemble SPOD (left) and Welch-based SPOD (right). The corresponding eigenvalues are marked by yellow crosses in the top ($\Delta f = 25\,$Hz) and bottom ($\Delta f = 100\,$Hz) panels of Fig.~\ref{fig:cavity1}.}
    \label{fig:cavity3}
\end{figure}

Having discussed the eigenvalue spectra and their convergence behavior, we now focus on the mode shapes. The modes corresponding to eigenvalues marked with yellow arrows and circles in Fig.~\ref{fig:cavity1} are shown in Figs.~\ref{fig:cavity2} and \ref{fig:cavity3}.  \rev{The plots compare bSPOD modes computed from $N_f$ neighboring Fourier modes with Welch-SPOD modes evaluated at the lower-bound Fourier-bin frequency of the corresponding interval. Comparisons with modes evaluated at the upper frequency bound lead to the same conclusions and are therefore omitted for brevity.} 

We begin with a tonal component at 1540~Hz for the fine frequency-resolution case, $\Delta f=25$~Hz, corresponding to the eigenvalue spectrum shown in the top panel of Fig.~\ref{fig:cavity1}. 
Figure~\ref{fig:cavity2} shows the real part of the vertical-velocity mode, with the bSPOD result on the left and the Welch-SPOD result on the right. As expected, the two mode shapes are nearly identical, since they are estimated from CSD matrices of equal size \rev{and are associated with the same frequency range} (1525--1550~Hz). Interestingly, both modes appear well converged despite the high variance in the spectra, see top panel in Fig.~\ref{fig:cavity1}.
This is because the tonal component carries most of the power in that frequency bin, producing a sharply defined contribution to the cross-spectral density and, in turn, a clearly identifiable eigenvector, even at high frequency resolution.

A different behavior is observed for broadband dynamics of lower power. The top row of Fig.~\ref{fig:cavity3} shows the real part of the vertical-velocity mode for SPOD and bSPOD at the same frequency resolution, in the 425 – 450~Hz band where the leading mode is broadband (see the top panel of Fig.~\ref{fig:cavity1}).
Here, the high frequency resolution and consequently small number of Fourier modes ($N_f = N_b =25$) contributing to the \rev{estimation of the} CSD matrix lead to a noisy and poorly converged mode estimates. \rev{Importantly, the SPOD and bSPOD modes exhibit a similar level of noise and overall quality, indicating that neither method provides a clear advantage in this case.} 
The bottom row of Fig.~\ref{fig:cavity3} shows the corresponding mode shapes for a lower frequency resolution, $\Delta f = 100$~Hz, such that the mode represents the leading dynamics in the 400–500~Hz band. 
For both methods, the \rev{increase in} frequency \rev{spacing} increases the size of the eigenvalue problem ($N_f = N_b =100$), implying that more orthogonal modes contribute to the representation of the dynamics within the band. 
As commonly observed for SPOD this reduces the noise level seen in Fig.~\ref{fig:cavity3} equally for the Welch and the band-ensemble approach. 

It is noted, that this improvement is a result of the dominant flow dynamics not varying substantially in the considered frequency bands;
for a finite dataset, increasing the \rev{number of realizations}, either by adding additional blocks (SPOD) or consecutive Fourier frequencies (bSPOD), necessarily \rev{increases the effective frequency range contributing to each spectral estimate.}
If the mode shape varies strongly with frequency, for example due to strong dispersion of the dominant wavelength, \rev{increasing the effective spacing} mixes dissimilar dynamics and can offset the benefit of added \rev{ realizations}. In this case improved convergence can only be achieved by considering a longer time series. \\

\begin{figure}
    \centering
    \includegraphics[width=0.49\linewidth]{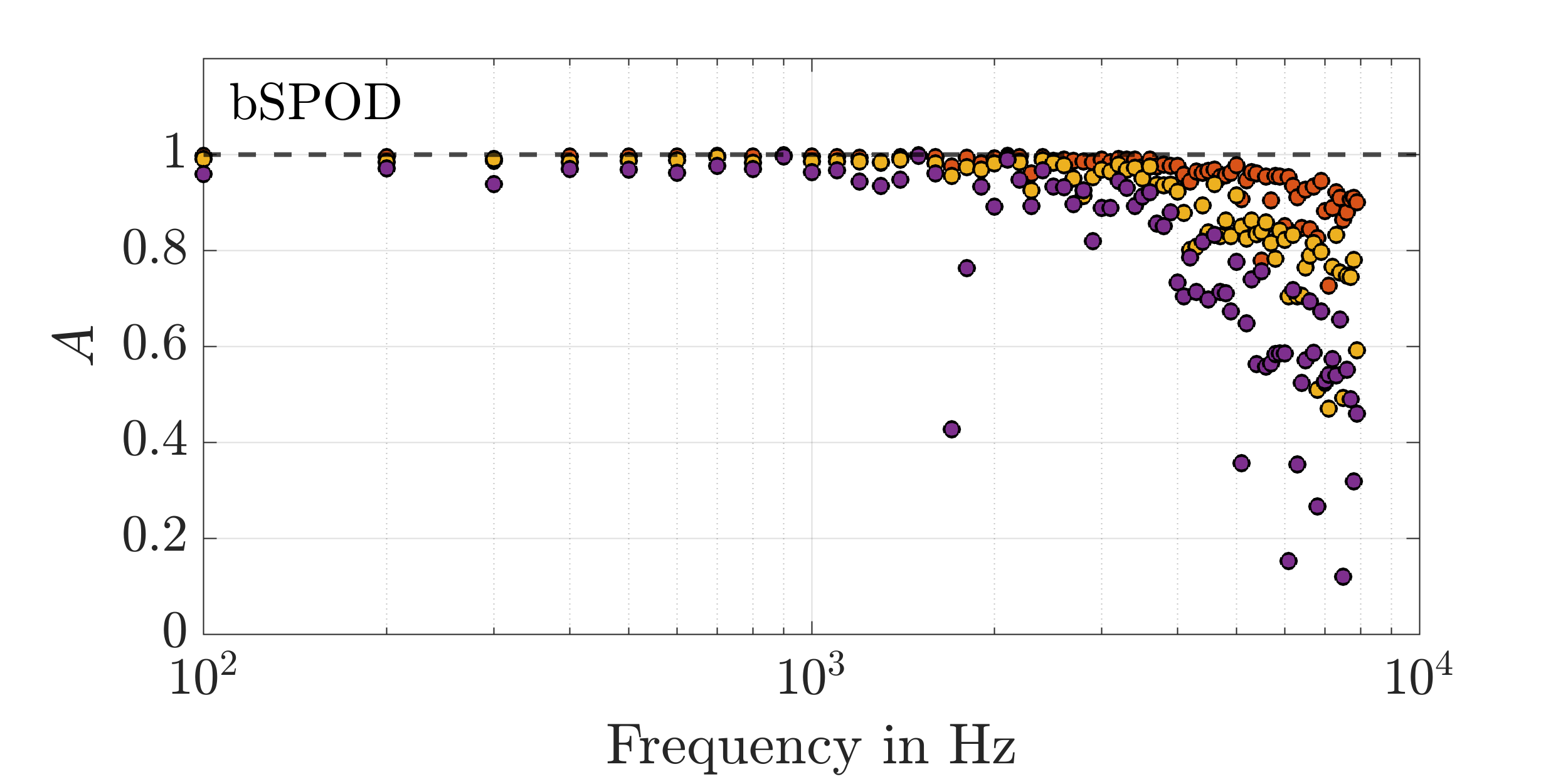}
     \includegraphics[width=0.49\linewidth]{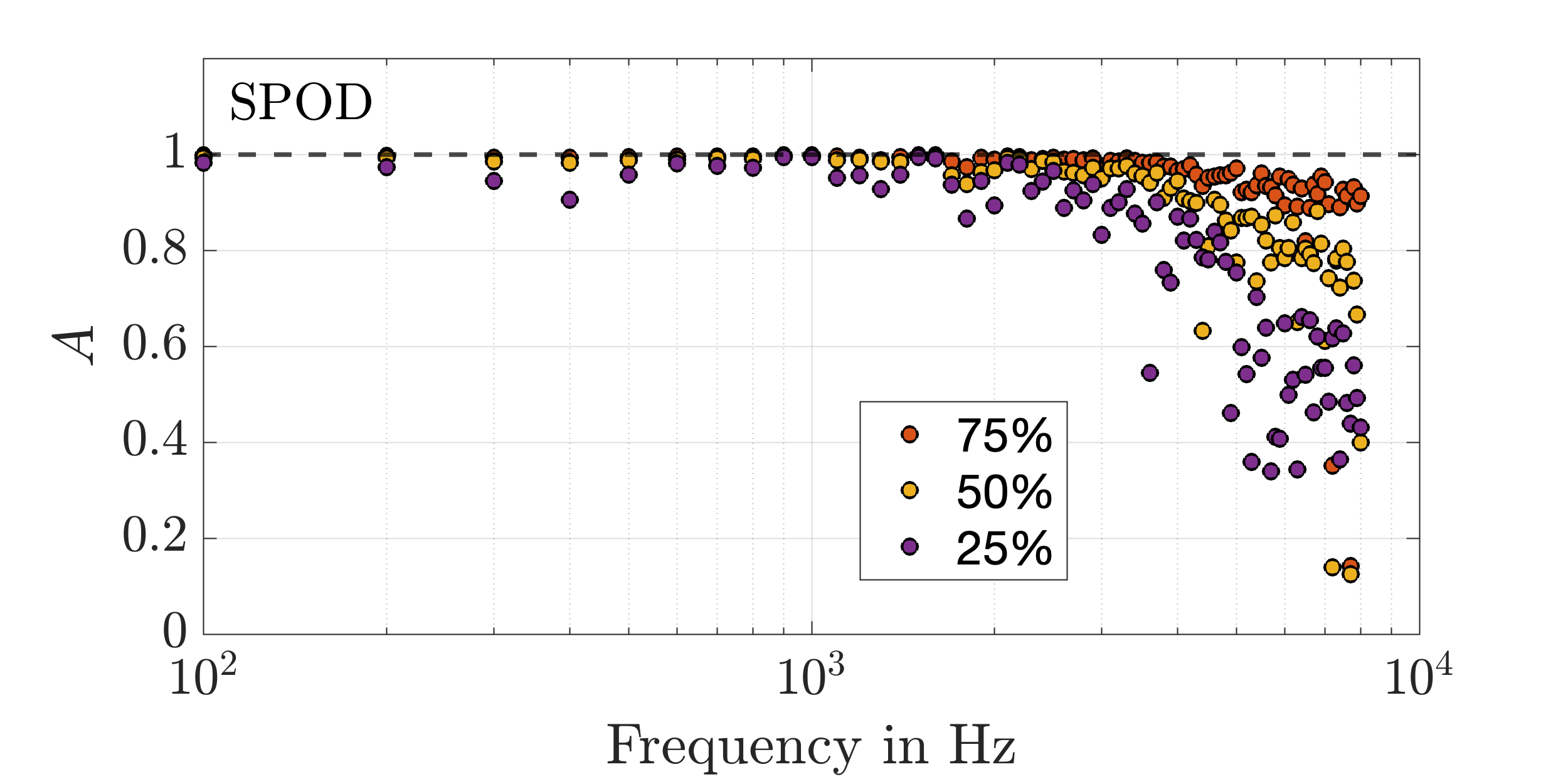}
    \caption{Mode convergence of the dominant SPOD eigenvalue assessed by reducing the number of snapshots to 75, 50, and 25\% of the full time record. Results are shown for band-ensemble SPOD (left) and Welch-based SPOD (right). Convergence is quantified by keeping the frequency resolution fixed and aligning the leading SPOD mode shape computed from reduced time sequences with the corresponding mode obtained from the full time record (see Eq.~\eqref{eq:alignment}).}
    \label{fig:cavity4}
\end{figure}

Finally, Fig.~\ref{fig:cavity4} compares the mode convergence behavior of SPOD and bSPOD with respect to the amount of available data. Therefore the dataset was progressively shortened from 100\% to 25\% in steps of 25\%, while keeping the frequency \rev{spacing} constant at $\Delta f = 100$~Hz. This implies that the number of modes representing the dynamics within each \rev{effective} frequency \rev{band} decreases proportionally with data reduction. 

Figure~\ref{fig:cavity4} shows the alignment between the leading modes (largest eigenvalue) obtained from the reduced, $\boldsymbol{\phi}_{\mathrm{red}}$, and full datasets, $\boldsymbol{\phi}_{\mathrm{ref}}$, defined as
\begin{equation}\label{eq:alignment}
A = \frac{|\boldsymbol{\phi}_{\mathrm{red}}^*  \boldsymbol{\phi}_{\mathrm{ref}}|}
{\|\boldsymbol{\phi}_{\mathrm{red}}\|\,
 \|\boldsymbol{\phi}_{\mathrm{ref}}\|},
\end{equation}
which is a commonly used measure of modal similarity~\citep{Cavalieri2013}. An alignment value of 1 means that the modes are fully aligned, while a value of 0 indicates that the mode shapes are orthogonal.
As expected, Fig.~\ref{fig:cavity4} shows that the alignment increases with the amount of data considered for both methods. Although the exact values differ for individual frequencies, SPOD and bSPOD exhibit comparable convergence behavior. 

We conclude that, under equivalent conditions $N_b=N_f$, \rev{corresponding to the same effective frequency bandwidth}, the dynamics \rev{are} represented by the same \rev{number} of POD modes, \rev{and} SPOD and bSPOD yield similar mode shapes with comparable convergence characteristics.

\subsection{Frequency attribution}
So far we have seen that bSPOD achieves mode convergence comparable to Welch-based SPOD while reducing spectral leakage and improving the frequency estimates of the eigenvalues via the frequency-attribution step. This advantage arises because bSPOD forms the CSD estimate from consecutive Fourier modes, so the method-of-snapshots eigenvalues (the expansion coefficients $\boldsymbol{\Theta}$, Eq.~\eqref{eq:bSPOD_m_of_snaps}) quantify how much each discrete Fourier mode contributes to a given bSPOD mode (Eq.~\eqref{eq:bSPOD_n_of_snaps2}). After normalization, the coefficients act as frequency weights which bSPOD uses to infer an in-band frequency estimate for each mode (Eq.~\ref{eq:weighted_freqs}).
In this section we elaborate on the frequency attribution step for the broadband cavity flow.

\begin{figure}
    \centering
    \includegraphics[width=0.99\linewidth]{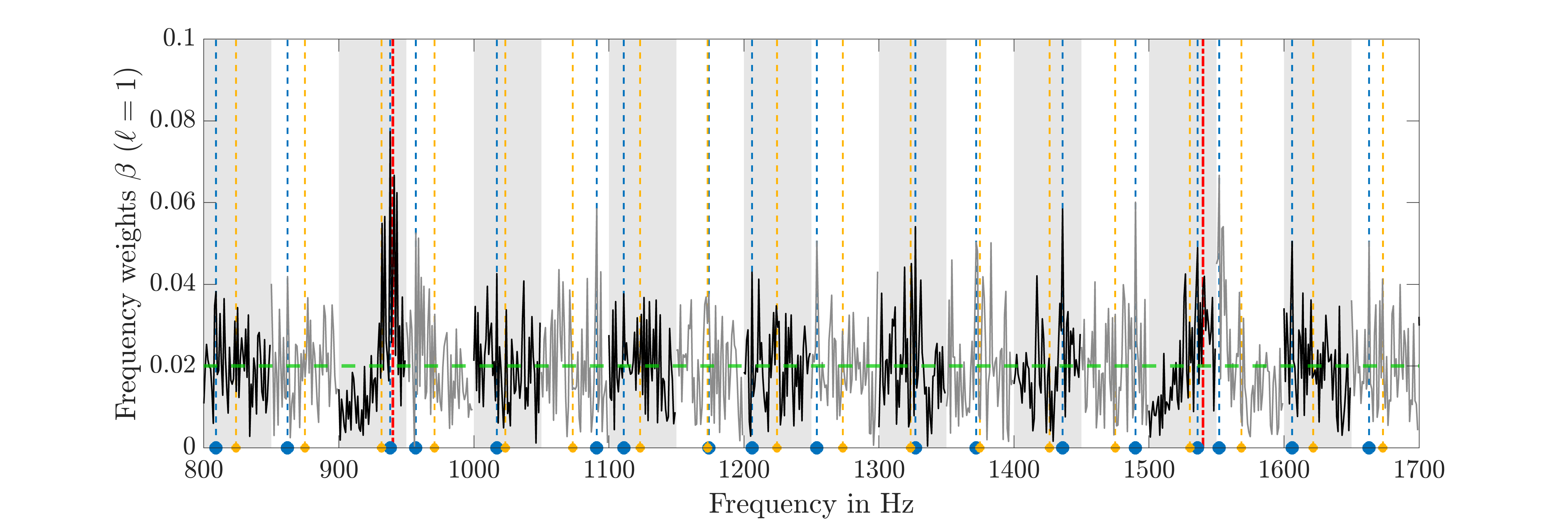}
    \includegraphics[width=0.99\linewidth]{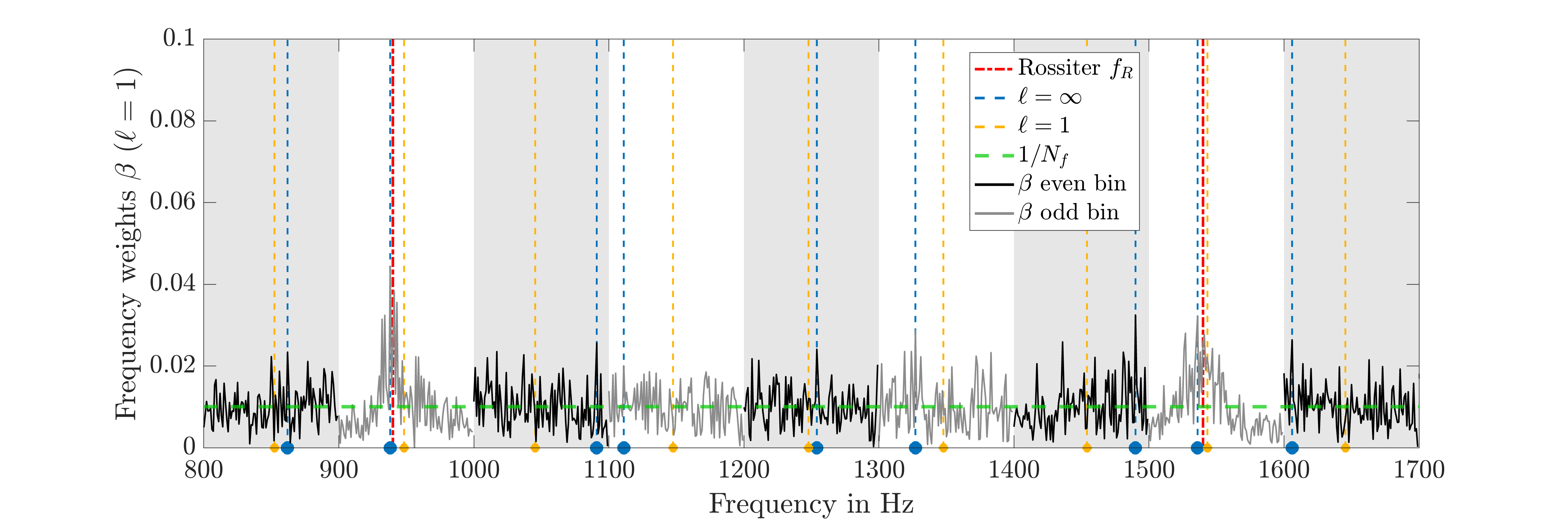}
    \caption{Frequency weights $\boldsymbol{\beta}$ of the leading band-ensemble mode versus frequency for $\ell=1$ and $\Delta f=50\,$Hz (top) and $100\,$Hz (bottom), see Eq.~\eqref{eq:weight}. The alternating grey–white background and alternating black/grey curves indicate the frequency bands used in bSPOD. Vertical blue and yellow lines mark the bSPOD frequency estimates obtained with $\ell=1$ and $\ell=\infty$, respectively. Red vertical lines denote the Rossiter-mode frequencies (see Fig.~\ref{fig:cavity2}).}
    \label{fig:cavity5}
\end{figure}

Figure~\ref{fig:cavity5} shows the frequency weights $\boldsymbol{\beta}$ of the leading cavity-flow mode for $\ell=1$ and two resolutions, $\Delta f=50$ and $\Delta f=100$~Hz (top and bottom). The corresponding eigenvalue spectra are shown above in the middle and bottom panels of Fig.~\ref{fig:cavity1}. Weights are plotted against the discrete Fourier frequencies contributing to each band between 800 and 1700~Hz; even and odd bands are shown in black and gray to emphasize the band-wise definition. In most bands, the weights fluctuate around the mean $1/N_f$ (green horizontal line) without a distinct pattern. In bands containing a tone, marked by red vertical lines, however, weights near the tone are elevated and contribute more strongly to the bSPOD mode, reflecting the tone’s higher power and clear separability from the surrounding broadband dynamics.

The blue dashed lines mark bSPOD frequency estimates from the attribution step using $\ell=\infty$ (Eq.~\eqref{eq:weighted_freqs}), selecting the frequency from the single most contributing Fourier mode per band. In tonal bands these lie very close to the Rossiter frequencies (red).
At $\Delta f=50\,$Hz, both shown Rossiter frequencies lie near an upper band limit. The algorithm accounts for this and assigns frequencies (blue) close to the tone (red) not only to the band containing the tone but also to the adjacent band by placing an attributed frequency at the lower edge of the neighboring band. This yields spikier modal power densities (Fig.~\ref{fig:cavity3}, middle) than those obtained with Welch-based SPOD, which uses a fixed frequency grid. 

The yellow dashed lines in Fig.~\ref{fig:cavity5} show the bSPOD frequency estimates obtained with the shown weights ($\ell=1$). These are less accurate at representing the tonal frequencies than the $\ell=\infty$ estimates, but they still provide better frequency estimates than a fixed choice such as the band-center frequency. In broadband bands without tones, the $\ell=1$ estimates tend to select frequencies closer to the band center, consistent with power being distributed across the band. In these broadband bands, $\ell=\infty$ can yield frequency estimates near the band edges even when the contributions are nearly uniform. Hence, it is desirable to rely on $\ell=\infty$ for tonal bands and $\ell=1$ for broadband bands; such a switch could be triggered by the local variation of the eigenvalue power density with frequency. In all cases, the normalization of the frequency weights, Eq.~\eqref{eq:weight}, ensures that the estimated frequency remains bounded within the respective band, so neither $\ell=1$ nor $\ell=\infty$ yields an out-of-band frequency estimate. Consequently, in contrast to Welch-based SPOD, which relies on a fixed frequency grid, the frequency estimation can only improve.

\subsection{Overlap in band-ensemble SPOD}\label{sec:overlap}
For Welch-based SPOD, it is common to overlap the time segments to increase the number of Fourier estimates from a finite time signal. The same principle can be used in bSPOD by shifting the band-ensemble window by less than $N_f$, so that consecutive frequency-blocks overlap. This produces additional estimates and \rev{thereby refines the} frequency \rev{grid} of the \rev{modal} power spectrum.

When applying overlap, however, two points must be considered. First, although the bSPOD modes are defined \rev{with} the \rev{frequency spacing} associated with the chosen band length $N_f$, the effective \rev{spacing} for the power estimate changes with overlapping because Fourier modes are reused across overlapping blocks. In practice, the overlap factor resulting in increased frequency resolution must be considered when integrating the power density over frequency to recover the full power of the signal. 

Second, in broadband signals containing tonal components, overlap can artificially spread the energy of a tone across several \rev{bSPOD} bins if all modes are simply assigned to the band-center frequency ($\ell = 0$). For PSD estimates using frequency smoothing, this produces block-shaped jumps in the spectrum, likely a reason why PSDs are more commonly estimated using Welch segmentation rather than frequency smoothing. In bSPOD, this issue is avoided by the data-driven frequency attribution step, enabled by the POD expansion coefficients. It ensures that each bSPOD mode is linked to its most contributing frequencies and leads to the tonal power being correctly concentrated. 

\begin{figure}
    \centering
    \includegraphics[width=0.99\linewidth]{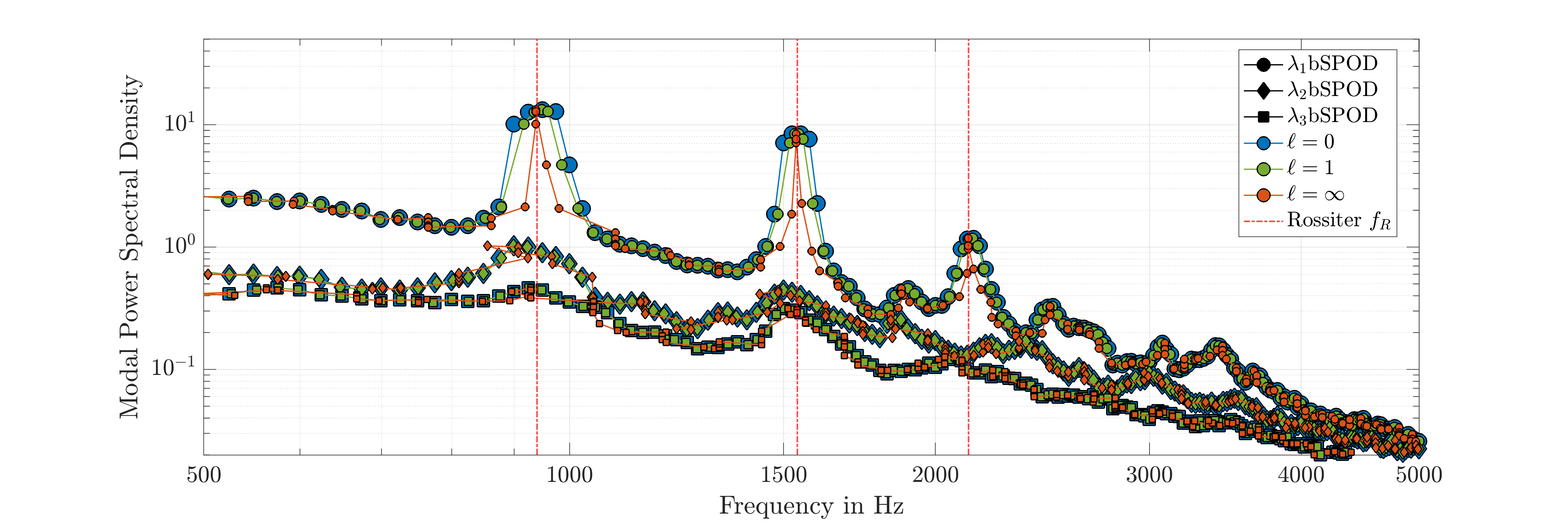}
    \caption{Leading three bSPOD eigenvalues versus frequency for the cavity-flow data, computed with $N_f=100$ ($\Delta f=100\,$Hz) and 75\% band overlap. Eigenvalues are shown for a fixed frequency grid ($\ell=0$, assigning each eigenvalue to the band center) and for two data-driven frequency attributions, $\ell=1$ and $\ell=\infty$ (see Eq.~\eqref{eq:weight}).}
    \label{fig:cavity6}
\end{figure}

To illustrate this behavior, Fig.~\ref{fig:cavity6} shows the leading three bSPOD eigenvalues of the cavity-flow data for $N_f=100$ ($\Delta f=100~\mathrm{Hz}$; compare with the bottom panel of Fig.~\ref{fig:cavity1}) and a band overlap of 75\%. Compared to the case without overlap, the frequency resolution is increased by a factor of four. It should be noted that adjacent modes partially share the same Fourier components, leading to statistically correlated estimates. Results are shown for frequency attribution using $\ell=\infty$ and $\ell=1$, as well as for the case without attribution, where each bSPOD mode is assigned the band-center frequency ($\ell=0$).  

As expected, without frequency attribution ($\ell=0$) the spectrum resembles classical frequency smoothing: tonal peaks appear as block-wise elevations. With $\ell=\infty$, the tonal modes are attributed frequencies close to the Rossiter frequencies, thereby preserving sharp tonal peaks in the spectrum despite the band overlap. In the broadband parts of the spectrum, however, $\ell=\infty$ can yield uneven or discontinuous frequency trends, including small jumps, which are particularly apparent for the first subleading mode in Fig.~\ref{fig:cavity6}. Using $\ell=1$ produces a smooth and well-resolved broadband spectrum, but yields less sharply localized tonal peaks than $\ell=\infty$, although still improved compared to $\ell=0$. These observations suggest a hybrid strategy in which $\ell=1$ is used in broadband bands and $\ell=\infty$ in bands containing tones. As mentioned above, such a switch could be triggered, for example, by detecting strong local variations of the eigenvalue power density with frequency.

We conclude that, in bSPOD, introducing overlap between neighboring frequency bands can increase the effective frequency resolution of the modal spectrum without smearing tonal components across adjacent bins, provided that an appropriate frequency attribution is used. In Welch-based SPOD, overlap is typically introduced to increase the number of blocks of fixed length. Alternatively, overlap in Welch-based SPOD can be interpreted as allowing for increased block length at a fixed number of blocks, leading to a comparable effect on frequency resolution. The key advantage of bSPOD is that overlap is applied after the Fourier transform and can be varied across the spectrum, whereas SPOD based on time segmentation relies on a fixed frequency grid and does not offer this flexibility.

\section{Conclusions}
This work presents a method for computing SPOD modes of statistically stationary data. The proposed approach, termed band-ensemble SPOD, is inspired by frequency smoothing, a method to reduce estimation variance in power spectral density estimates. Here, this methodology is extended to SPOD, enabling the estimation of spatial modes together with their modal power spectral densities.
Unlike the more commonly applied Welch-based SPOD formulation, which estimates the cross-spectral density matrix from Fourier modes defined on the same frequency bin obtained through temporal segmentation, bSPOD approximates the cross-spectral density matrix, and thus the SPOD modes, from \rev{narrowly spaced}, neighboring Fourier coefficients. These coefficients are obtained from a single Fourier transform of the full time series, without segmentation.

Compared with Welch-based SPOD, bSPOD offers several advantages. First, spectral leakage is naturally reduced by avoiding time segmentation. Second, the contribution of individual Fourier coefficients to each bSPOD mode can be tracked through the expansion coefficients, allowing in-band frequency information to be retained. This information can be exploited to obtain data-driven frequency estimates for individual bSPOD modes. By combining these properties, bSPOD reduces estimation variance while maintaining low bias for tonal components, providing a favorable balance in the bias–variance trade-off relative to the Welch-based formulation. As a result, bSPOD is particularly well suited for broadband–tonal flows, where variance reduction without compromising the accuracy of tonal eigenvalue estimates is desirable.

The bSPOD formulation further allows this trade-off to be adjusted by varying the filter length with frequency. This enables modal decompositions over variable frequency bands with a frequency-dependent number of modes. Such flexibility makes bSPOD well suited for adaptive strategies that select the number of modes based on convergence criteria, as suggested by~\citet{yeung2024}. Given these properties and a computational cost comparable to Welch-based SPOD, bSPOD provides a practical improvement for spectral modal analysis of turbulent flows.

\begin{acknowledgments}
This work was supported by a postdoc
fellowship of the German Academic Exchange Service (DAAD).
\end{acknowledgments}

\appendix


\bibliography{apssamp}

\end{document}